\documentclass[]{article}

\usepackage[a4paper, total={6in, 8in}]{geometry}
\usepackage[]{algorithm2e}
\usepackage{amsmath}
\usepackage{amsfonts}
\usepackage{amsthm}
\usepackage{authblk}
\usepackage[ngerman, UKenglish]{babel}
\usepackage{bbm}
\usepackage{bigstrut}
\usepackage{caption}
\usepackage{cases}
\usepackage{comment}
\usepackage{float}
\usepackage[acronym]{glossaries}
\usepackage{graphicx, subfigure}
\usepackage[colorlinks=true,bookmarksopen,bookmarksnumbered,linkcolor=black,citecolor=black,urlcolor=black]{hyperref}
\usepackage[utf8]{inputenc}
\usepackage{moreverb}
\usepackage{multirow}
\usepackage{pgfplots}
\pgfplotsset{compat=1.3}
\usepackage{placeins}
\usepackage{tabularx}
\usepackage{tikz}

\makeglossaries
\DeclareMathAlphabet\mathbfcal{OMS}{cmsy}{b}{n}
\newacronym{em}{EM}{electromagnetic}
\newacronym{emf}{EMF}{electromagnetic field}
\newacronym{fem}{FEM}{finite element method}
\newacronym{gpc}{gPC}{generalized polynomial chaos}
\newacronym{hf}{HF}{high-frequency}
\newacronym{mc}{MC}{Monte Carlo}
\newacronym{pde}{PDE}{partial differential equation}
\newacronym{pdf}{PDF}{probability density function}
\newacronym{qoi}{QoI}{quantity of interest}
\newacronym{rms}{RMS}{root-mean-square}
\newacronym{rmse}{RMSE}{root-mean-square error}
\newacronym{rv}{RV}{random variable}
\newacronym{uq}{UQ}{uncertainty quantification}

\title{Assessing the Performance of Leja and Clenshaw-Curtis Collocation for Computational Electromagnetics with Random Input Data}
\author[1,2]{Dimitrios Loukrezis}
\author[3]{Ulrich R\"omer}
\author[1,2]{Herbert De Gersem}

\affil[1]{{\small Institut f\"ur Theorie Elektromagnetischer Felder, Technische Universit\"at Darmstadt, \newline Schlossgartenstra\ss e 8, 64289 Darmstadt, Germany}}
\affil[2]{{\small Centre for Computational Engineering, Technische Universit\"at Darmstadt,  \newline Dolivostra\ss e 15, 64293 Darmstadt, Germany}}
\affil[3]{{\small Institut f\"ur Dynamik und Schwingungen, Technische Universit\"at Braunschweig, \newline Schleinitzstra\ss e 20, 38106 Braunschweig, Germany}}

\DeclareMathOperator*{\argmax}{argmax}
\date{}

\begin{document}

\maketitle

\providecommand{\keywords}[1]{\textbf{\textit{keywords--}} #1}

\begin{abstract}
We consider the problem of quantifying uncertainty regarding the output of an electromagnetic field problem in the presence of a large number of uncertain input  parameters.
In order to reduce the growth in complexity with the number of dimensions, we employ a dimension-adaptive stochastic collocation method based on nested univariate nodes.
We examine the accuracy and performance of collocation schemes based on Clenshaw-Curtis and Leja rules, for the cases of uniform and bounded, non-uniform random inputs, respectively.
Based on numerical experiments with an academic electromagnetic field model, we compare the two rules in both the univariate and multivariate case and for both quadrature and interpolation purposes. 
Results for a real-world electromagnetic field application featuring high-dimensional input uncertainty are also presented.

\end{abstract}

\section{Introduction}
More often than ever before, the design phase of electric and electronic devices, e.g. waveguides or accelerator magnets, incorporates  parameter studies in order to predict the device's behavior under uncertainty.
This uncertainty, e.g. with respect to the device's geometry or material properties, often stems from tolerances during the manufacturing process.
As part of those \gls{uq} studies, one typically investigates a specific output of the device, called the \gls{qoi}, and tries to estimate statistical moments or sensitivities, with the goal of reducing the risk of malfunction, misfire or other type of failure.

Most commonly, \gls{uq} studies rely on sampling methods.
\gls{mc} sampling \cite{caflisch1998} converges irrespective of the number of \glspl{rv} or the regularity of the given problem, albeit with a slow convergence rate of $\mathcal{O}\left(M^{-0.5}\right)$ in the mean-square-error sense, where $M$ denotes the number of samples, equivalently, costs.
Improved cost-error ratios can be achieved with multilevel \gls{mc} \cite{giles2015} methods.
Spectral \gls{uq} approaches \cite{Ghanem1991, xiu2010} converge much faster, exponentially in the best case, for a small to moderate number of random inputs and smooth input-to-output map. 
Typical methods of this type are stochastic Galerkin \cite{journals/siamnum/BabuskaTZ04, Ghanem1991, MatthiesGalerkin}, stochastic collocation \cite{journals/siamrev/BabuskaNT10, journals/adcm/BarthelmannNR00, bungartz_griebel_2004, journals/siamsc/XiuH05}, or point collocation \cite{journals/jcphy/BlatmanS11, journals/siamsc/MiglioratiNST13, journals/focm/MiglioratiNST14} methods.

The stochastic Galerkin method is often labeled as ``intrusive'', due to the fact that dedicated solvers have to be developed in order to tackle the stochastic problem at hand.
The additional programming effort is usually regarded as a major disadvantage, especially in the case of complex computational models whose software and underlying solvers are difficult to be accessed, modified or otherwise manipulated.
Therefore, and despite the fact that stochastic Galerkin methods have nice properties for error analysis and estimation, collocation methods are generally preferred, as they allow for a non-intrusive, black-box use of the original computational models.
It must be noted that the separation of methods into intrusive and non-intrusive is an ongoing topic of discussion, see e.g. \cite{journals/siamsc/GiraldiLLMN14}. 
In the context of the present paper we shall retain the usual distinction.
Comparisons between stochastic and point collocation methods, see e.g. \cite{Eldred2009comparison}, indicate that the former tends to provide superior accuracies and convergence rates for smooth \glspl{qoi}. 
However, since these approaches differ significantly, a fair comparison between the two is still an open research topic, as also indicated in \cite{journals/siamsc/MiglioratiNST13}.

A common bottleneck of all aforementioned methods is the so-called ``curse of dimensionality'' \cite{Bellman:57}, i.e. convergence rates deteriorate and computational costs increase with the number of considered input parameters, by definition, exponentially.
As a possible remedy, state-of-the-art methods employ sparse, adaptively constructed polynomial approximations, see e.g. \cite{journals/focm/ChkifaCS14, journals/siamsc/NarayanJ14, journals/siamnum/NobileTW08a, SchiechePhD} for adaptive stochastic collocation methods, \cite{journals/jcphy/BlatmanS11, conf/enumath/Migliorati13} for adaptive point collocation methods, and \cite{eigel2014} for an adaptive stochastic Galerkin method.
While generally not free of the curse of dimensionality, adaptive methods exploit possible anisotropies among the input parameters  regarding their impact upon the \gls{qoi}.
Assuming that such anisotropies exist, adaptivity may enable studies with a comparably large number of input parameters.
More recently, tensor decompositions (see \cite{hackbusch2014} and the references therein) have been used to exploit possible low-rank structures of parametric problems in order to tackle the curse of dimensionality.
In several cases, again relying on high regularity, superior asymptotic convergence rates have been obtained compared to sparse grid methods \cite{UschmajewPhD}.
However, comparisons between these methods remains an active field of research, as break-evens have not yet been fully determined.

Here we will only consider stochastic collocation methods, in which case, dimension-adaptive algorithms \cite{journals/computing/GerstnerG03, journals/toms/KlimkeW05} constitute the current state-of-the-art.
In the search for an acceptable compromise between computational work and approximation accuracy, such approaches are receiving increasing attention in uncertainty quantification.
Dimension-adaptive methods are based on nested univariate collocation points, e.g. Clenshaw-Curtis and Genz-Keister nodes are typical choices for uniform and normal input distributions, respectively.
The extension of dimension-adaptive schemes to cases where the input distributions do not fall into the two aforementioned categories is desirable and an active field of interest, as well as one of the main considerations of the present paper.

In this work we consider univariate and medium to high-dimensional multivariate \gls{uq}, in the context of \gls{emf} problems with random inputs.
The probability distributions of the inputs are assumed to be bounded, but not necessarily uniform, e.g. beta distributions are considered in this work.
The stochastic collocation method is used for the \gls{uq} studies.
When multiple inputs are considered, we employ a dimension-adaptive algorithm based on nested univariate collocation points.
We investigate the performance of the method for different choices of nested collocation points, in particular provided from either Clenshaw-Curtis \cite{clenshaw1960} or Leja rules \cite{leja1957}.
In the case of non-uniform input distributions, we use weighted Leja rules based on \cite{journals/siamsc/NarayanJ14}, while the Clenshaw-Curtis rules are modified as in \cite{sommariva2013fast}.
For the case of uniform inputs only and in a purely mathematical context, comparisons between the two rules can be found in \cite{journals/focm/ChkifaCS14, journals/siamsc/NarayanJ14, Nobile2015}.
We are unaware of such comparisons for the case of bounded, non-uniform inputs, such as the ones considered here. 
The available literature also lacks works considering concrete engineering applications, such as the \gls{emf} problems presented in this work.

The rest of the paper is organized as follows.
In Section~\ref{sec:setting} we offer a general description of the \gls{uq} problem at hand.
In Section~\ref{sec:sgsc} we describe the \gls{uq} method of choice, namely the stochastic collocation method.
The univariate collocation is presented in Section~\ref{sec:univarcolloc}, while multivariate collocation schemes are presented in Sections~\ref{sec:tpcolloc} and \ref{sec:sparsegrids} for the cases of tensor grids and sparse grids, respectively.
The latter are further separated into isotropic and adaptive-anisotropic sparse grids, respectively discussed in Sections~\ref{sec:iso_sg} and \ref{sec:adapt_sg}.
Post-processing and collocation-based quadrature schemes are presented in Section~\ref{sec:quad_postproc}.
The two choices of collocation points considered in this work are presented in Section~\ref{sec:colloc_points}.
A number of numerical experiments are given in Section~\ref{sec:numexp}. 
An analytical, academic \gls{emf} model is considered in Section~\ref{sec:waveguide}.
Results for a real-world application are available in Section~\ref{sec:sterngerlach}.

\section{Problem Setting}
\label{sec:setting}
Let $\left(\Theta, \Sigma, P\right)$ be a probability space, $\theta \in \Theta$ a random event and $\mathbf{Y} = \left(Y_1, Y_2, \dots, Y_N\right)$ a vector of $N$ independent \glspl{rv} defined on $\left(\Theta, \Sigma, P\right)$.
We denote with $\mathbf{y} = \mathbf{Y}\left(\theta\right)$ a random realization of the input \glspl{rv} and with $\rho$ their joint \gls{pdf}, such that $\mathbf{Y} : \Theta \rightarrow \Xi \subset \mathbb{R}^N$  and $\rho : \Xi \rightarrow \mathbb{R}_+$, where $\mathbb{R}_+ = \left\{y \in \mathbb{R} : y > 0\right\}$.
In the context of the present work, $\Xi$ is an $N$-dimensional hyper-rectangle $\Xi = \Xi_1 \times \Xi_2 \times \cdots \times \Xi_N$. 
A univariate \gls{pdf} $\rho_n\left(y_n\right)$ corresponds to each $Y_n$, $n=1,2,\dots,N$.
Due to the statistical independence of the \glspl{rv}, the joint \gls{pdf} is given by
\begin{equation}
\label{eq:independence}
 \rho\left(\mathbf{y}\right) = \prod_{n=1}^N \rho_n\left(y_n\right).
\end{equation}

We now assume a \gls{pde} given in the general form
\begin{equation}
\label{eq:general}
 \mathcal{D}(u, \mathbf{y}) = 0,
\end{equation}
where $u = u\left(\mathbf{y}\right)$ is the solution of \eqref{eq:general} and $\mathbf{y} \in \mathbb{R}^N$ is a parameter vector.
We consider the \gls{pde} to be well posed for all $\mathbf{y} \in \Xi \subset \mathbb{R}^N$.
Specifics regarding the \glspl{pde} of the mathematical models of the here considered \gls{emf} problems are discussed in detail in Sections~\ref{sec:waveguide} and \ref{sec:sterngerlach}.
We assume a model output which is given as a functional $q\left(u\left(\mathbf{y}\right)\right)$, commonly called the \gls{qoi}. 
For simplicity, we denote the parameter-dependent \gls{qoi} with $q\left(\mathbf{y}\right)$, and assume that $q(\mathbf{y}) \in \mathbb{R}$, however, complex and vector-valued \glspl{qoi} may also be considered.
Preferably, the dependence of the \gls{qoi} on the input \glspl{rv}, as given by $q$, is smooth, ideally analytic.

In the case where $\mathbf{y} = \mathbf{Y}\left(\theta \right)$, i.e. the parameter vector constitutes a realization of a random vector, the \gls{qoi} is also a \gls{rv} given by $q\left(\mathbf{Y}\right)$.
In other words, the input uncertainty propagates through the (deterministic) model and renders the \gls{qoi} uncertain as well.
The task at hand is to quantify the uncertainty of the now random \gls{qoi}, e.g. by computing statistical moments, sensitivities with respect to the random inputs, event probabilities, etc.
A typical \gls{uq} task, such as the aforementioned ones, can be written in the general form
\begin{equation}
\label{eq:uq_task}
 \mathbb{E}\left[\phi\left(q\right)\right] = \int_{\Xi} \phi\left(q\left(\mathbf{y}\right)\right) \rho(\mathbf{y}) \mathrm{d}\mathbf{y},
\end{equation}
where $\phi$ denotes a functional corresponding to the sought statistical measure. 
For example, $\phi(q) = q$ in the case of the expected value $\mathbb{E}\left[q\right]$, or  $\phi(q) = \left(q - \mathbb{E}\left[q\right]\right)^2$ in the case of the variance $\mathbb{V}\left[q\right]$.

Assuming that $\phi$ is a continuous smooth function, the multivariate integral in \eqref{eq:uq_task} can be computed efficiently  with appropriate quadrature schemes, such that
\begin{equation}
 \label{eq:uq_task_quad}
 \mathbb{E}\left[\phi\left(q\left(\mathbf{y}\right)\right)\right] \approx \sum_{k=1}^K w^{(k)} \phi\left(q\left(\mathbf{y}^{(k)}\right)\right),
\end{equation}
where $\mathbf{y}^{(k)}$ and $w^{(k)}$ denote the $k$-th quadrature node and weight, respectively.
Alternatively, surrogate-based sampling methods can be employed, where the original model $q$ is substituted by an inexpensive surrogate model $\widetilde{q} \approx q$, assuming sufficient approximation accuracy. 
For example, the surrogate model may be an interpolation-based polynomial approximation 
\begin{equation}
 \label{eq:uq_task_interp} 
 q\left(\mathbf{y}\right) \approx \widetilde{q}\left(\mathbf{y}\right) = \sum_{k=1}^K q\left(\mathbf{y}^{(k)}\right) \Psi_k\left(\mathbf{y}\right),
\end{equation}
where $\mathbf{y}^{(k)}$ are interpolation nodes and $\Psi_k$ appropriate multivariate polynomials.
Both aforementioned approaches, i.e. quadrature and interpolation-based, can be efficiently implemented in the context of the stochastic collocation method, presented in Section \ref{sec:sgsc}.

\section{Stochastic Collocation}
\label{sec:sgsc}
In the stochastic collocation method, a polynomial approximation similar to \eqref{eq:uq_task_interp} is constructed by interpolating specific values of the \gls{qoi}.
Those values are provided by evaluating the \gls{qoi} on a set of realizations of the input \glspl{rv}, called collocation points.
We denote the set of collocation points with $Z$ and its cardinality with $\# Z$.
Since each evaluation requires a call to the original model, the computational cost of the method depends predominantly on $\# Z$.
The choice of collocation points depends on the \gls{pdf} $\rho$ which characterizes the input \glspl{rv}.
Quadrature rules for the approximation of \eqref{eq:uq_task} can be derived from the collocation, as will be shown in Section~\ref{sec:quad_postproc}.

\subsection{Univariate collocation}
\label{sec:univarcolloc}
Univariate interpolation rules are used as building blocks for stochastic collocation in multiple dimensions, respectively, multiple random parameters. 
Therefore, let us first consider the case of a single parameter, $Y$.

We introduce the non-negative integer $\ell \in \mathbb{N}_0$, called the interpolation level. 
The corresponding univariate grid of collocation points is denoted with $Z_{\ell}$. 
The number of univariate collocation points $\# Z_{\ell}$ is associated with the level $\ell$ through a monotonically increasing ``level-to-nodes'' function $m \: : \: \mathbb{N}_0 \rightarrow \mathbb{N}$, such that $\# Z_{\ell} = m\left(\ell\right)$, with $m\left(0\right) = 1$.
The choice of collocation points depends on the univariate \gls{pdf} $\rho(y)$.

The interpolation is based on Lagrange polynomials, defined by the univariate collocation points $Z_{\ell} = \Big\{y_{ \ell}^{(i)}\Big\}_{i=1}^{m\left(\ell\right)}$ as
\begin{align}
 \label{eq:lagrange1d}
 l_{\ell}^{(i)}\left(y\right) &= \prod_{\substack{k=
 1\\ k \neq i}}^{m\left(\ell\right)} \frac{y-y_{\ell}^{(k)}}{y_{\ell}^{(i)} - y_{\ell}^{(k)}}, \\
 l_{0}\left(y\right) &= 1. \nonumber
\end{align}
Denoting the univariate interpolation operator with $\mathcal{I}_{\ell}$, the interpolation reads
\begin{equation}
 \label{eq:interp_1d} 
 \mathcal{I}_{\ell} \left[q\right]\left(y\right) = \sum_{i=1}^{m\left(\ell\right)} q\left(y_{\ell}^{(i)}\right) l_{ \ell}^{(i)}\left(y\right).
\end{equation}

In the special case where the univariate grids are nested, i.e. $Z_{\ell-1} \subset Z_\ell$, the interpolation \eqref{eq:interp_1d} can be constructed in a hierarchical way, such that the \gls{qoi} must be evaluated only for the new collocation points $y_{\ell}^{(i)} \in Z_{\ell} \setminus Z_{\ell-1}$.
The hierarchical counterpart of \eqref{eq:interp_1d} reads
\begin{align}
 \label{eq:hinterp_1d} 
 \mathcal{I}_{\ell}\left[q\right]\left(y\right)
 &= \mathcal{I}_{\ell-1}\left[q\right]\left(y\right) 
 + \sum_{i : y_{\ell}^{(i)} \in Z_{\ell} \setminus Z_{\ell-1}}\Big( q\left(y_{\ell}^{(i)}\right) - \mathcal{I}_{\ell-1}\left[q\right]\left(y_{ \ell}^{(i)}\right)\Big) l_{\ell}^{(i)}\left(y\right) \\
 &= \mathcal{I}_{\ell-1}\left[q\right]\left(y\right) 
 + \sum_{i : y_{\ell}^{(i)} \in Z_{\ell} \setminus Z_{\ell-1}}s_{\ell}^{(i)} l_{\ell}^{(i)}\left(y\right), \nonumber
\end{align}
where $\mathcal{I}_{-1}$ is the null operator and the quantities 
\begin{equation}
 s_{\ell}^{(i)} = q\left(y_{\ell}^{(i)}\right) - \mathcal{I}_{\ell-1}\left[q\right]\left(y_{\ell}^{(i)}\right)
\end{equation}
are called hierarchical surpluses.
We further introduce the difference operator
\begin{equation}
 \Delta_{\ell} = \mathcal{I}_{\ell} - \mathcal{I}_{\ell-1},
\end{equation}
such that the interpolation operator in \eqref{eq:hinterp_1d} is given by
\begin{equation}
\label{eq:hinterp_1d_delta}
 \mathcal{I}_{\ell} = \sum_{k=0}^{\ell} \Delta_k.
\end{equation}
Nested univariate collocation rules are the key ingredients of adaptively constructed sparse grids, discussed in Section \ref{sec:adapt_sg}.

\subsection{Tensor-product collocation}
\label{sec:tpcolloc}
In its simplest form, multivariate collocation consists of tensor-product combinations of univariate interpolation rules.
We introduce the multi-index $\boldsymbol{\ell} = \left(\ell_1, \ell_2, \dots, \ell_N\right) \in {\mathbb{N}_0^N}$, which contains the interpolation level for each \gls{rv}.
Generally, the indices $\ell_1, \ell_2, \dots, \ell_N$ can have different values from one another.
The special case where $\ell_n = \ell$, $\forall n=1,2,\dots,N$, is called isotropic tensor-product collocation.

The multivariate collocation points are given as the tensor grid
\begin{align}
\label{eq:tensorgrid}
 Z_{\boldsymbol{\ell}} &= Z_{1,\ell_1} \times Z_{2,\ell_2} \times \cdots \times Z_{N,\ell_N}
 = \Big\{y_{1, \ell_1}^{(i_1)}\Big\}_{i_1=1}^{m_1\left(\ell_1\right)} \times \Big\{y_{2,\ell_2}^{(i_2)}\Big\}_{i_2=1}^{m_2\left(\ell_2\right)} \times \cdots \times \Big\{y_{N, \ell_N}^{(i_N)}\Big\}_{i_N=1}^{m_N\left(\ell_N\right)},
\end{align}
with cardinality $\#Z_{\boldsymbol{\ell}} = \# Z_{1,\ell_1} \# Z_{2,\ell_2} \cdots \# Z_{N,\ell_N}$.
Using the the multi-index $\mathbf{i} = \left(i_1, i_2, \dots, i_N\right)$, every multivariate collocation point $\mathbf{y}_{\boldsymbol{\ell}}^{(\mathbf{i})} \in Z_{\boldsymbol{\ell}}$ is given as $\mathbf{y}_{\boldsymbol{\ell}}^{(\mathbf{i})} = \left( y_{1,\ell_1}^{(i_1)}, y_{2,\ell_2}^{(i_2)}, \dots, y_{N, \ell_N}^{(i_N)}\right)$.
The corresponding multivariate Lagrange polynomials, $L^{(\mathbf{i})}_{\boldsymbol{\ell}}$ are defined as 
\begin{equation}
\label{eq:lagrangeNd}
 L^{(\mathbf{i})}_{\boldsymbol{\ell}}\left(\mathbf{y}\right) = \prod_{n=1}^N l_{n, \ell_n}^{(i_n)}\left(y_n\right).
\end{equation}
Denoting the multivariate Lagrange operator corresponding to the multi-index $\boldsymbol{\ell}$ with $\mathcal{I}_{\boldsymbol{\ell}}$, the tensor-product interpolation formula reads
\begin{align}
\label{eq:interp_tp}
 \mathcal{I}_{\boldsymbol{\ell}} \left[q\right]\left(\mathbf{y}\right) 
 &= \left(\mathcal{I}_{1,\ell_1} \otimes \mathcal{I}_{2,\ell_2} \otimes \cdots \otimes \mathcal{I}_{N,\ell_N}\right) \left[q\right]\left(\mathbf{y}\right) \nonumber \\ 
 &= \sum_{i_1=1}^{m_1\left(\ell_1\right)}\sum_{i_2=1}^{m_2\left(\ell_2\right)} \cdots \sum_{i_N=1}^{m_N\left(\ell_N\right)} q\left(y_{1,\ell_1}^{(i_1)}, y_{2,\ell_2}^{(i_2)}, \dots, y_{N,\ell_N}^{(i_N)}\right)
 \prod_{n=1}^N l_{n, \ell_n}^{(i_n)}\left(y_n\right) \\ \nonumber
 &= \sum_{\mathbf{i} : \mathbf{y}_{\boldsymbol{\ell}}^{(\mathbf{i})} \in Z_{\boldsymbol{\ell}}} q\left(\mathbf{y}_{\boldsymbol{\ell}}^{(\mathbf{i})}\right) L^{(\mathbf{i})}_{\boldsymbol{\ell}}\left(\mathbf{y}\right). \nonumber
\end{align}

While simple in its conception and construction, tensor-product stochastic collocation becomes intractable even for a moderate number of \glspl{rv}.
The curse of dimensionality is particularly evident in the case of isotropic tensor grids, where $\# Z_{\boldsymbol{\ell}} = m\left(\ell\right)^N$, i.e. the required computational work increases exponentially with respect to $N$.
Therefore, the use of tensor-product stochastic collocation is restricted to low-dimensional settings.

\subsection{Sparse Grids Collocation}
\label{sec:sparsegrids}
In high-dimensional settings, collocation on sparse grids is typically employed as a way to mitigate the computational cost of the full tensor-product collocation. 
Sparse grids were first introduced by Smolyak in \cite{Smolyak1963} and have been used in the context of the stochastic collocation method in a large number of works, see e.g. \cite{journals/siamrev/BabuskaNT10, journals/adcm/BarthelmannNR00, journals/focm/ChkifaCS14, journals/jcphy/JakemanW15, journals/siamsc/NarayanJ14, journals/siamnum/NobileTW08, journals/siamnum/NobileTW08a, SchiechePhD}.
Depending on the sparse grid's construction process, the collocation scheme will be called either isotropic, or adaptive-anisotropic.

In both cases, the collocation scheme is given as a linear combination of tensor-product interpolations, such that
\begin{equation}
 \label{eq:sg_general}
 \mathcal{I}_{\Lambda} \left[q\right]\left(\mathbf{y}\right) = \sum_{\boldsymbol{\ell} \in \Lambda} \Delta_{\boldsymbol{\ell}}\left[q\right]\left(\mathbf{y}\right),
\end{equation}
where $\Lambda$ is the set of all multi-indices $\boldsymbol{\ell}$ participating in the sum of \eqref{eq:sg_general}, fulfilling a sparsity constraint. 
The multivariate difference operator $\Delta_{\boldsymbol{\ell}}$ is given as
\begin{equation}
 \Delta_{\boldsymbol{\ell}} = \Delta_{1,\ell_1} \otimes \Delta_{2,\ell_2} \otimes \cdots \otimes \Delta_{N,\ell_N}.
\end{equation}
We enforce upon the set $\Lambda$ a monotonicity constraint, such that
\begin{equation}
\label{eq:monotonicity}
 \forall \boldsymbol{\ell} \in \Lambda \Rightarrow \boldsymbol{\ell}-\mathbf{e}_n \in \Lambda, \forall n = 1,2,\dots,N, \: \text{with} \: \ell_n > 0,
\end{equation}
where $\mathbf{e}_n = \left(\delta_{mn}\right)_{1 \leq m \leq N}$ is the $n$-th unit vector and $\delta_{mn}$ denotes the Kronecker delta.
Monotone sets, also known as downward-closed or lower sets, ensure that the telescopic property of the sum in \eqref{eq:sg_general} is preserved \cite{journals/computing/GerstnerG03}.
The corresponding grid of collocation nodes is given by
\begin{equation}
 Z_{\Lambda} = \bigcup_{\boldsymbol{\ell} \in \Lambda} Z_{\boldsymbol{\ell}}.
\end{equation}
We note that formula \eqref{eq:sg_general} is in general not interpolatory, except for the case of nested univariate collocation points \cite{journals/adcm/BarthelmannNR00}.

\subsubsection{Isotropic Sparse Grids}
\label{sec:iso_sg}
In the case of isotropic sparse grids \cite{journals/siamrev/BabuskaNT10, journals/adcm/BarthelmannNR00, bungartz_griebel_2004, journals/siamnum/NobileTW08}, we define the approximation level $L \in \mathbb{N}_0$ and enforce the restriction
\begin{equation}
\label{eq:isoconstraint}
 \left|\boldsymbol{\ell}\right| = \sum_{n=1}^N \ell_n \leq L,
\end{equation}
such that $\Lambda = \left\{\boldsymbol{\ell} : \left|\boldsymbol{\ell}\right| = \sum_{n=1}^N \ell_n \leq L\right\}$.
The term ``isotropic'' refers to the univariate interpolation rules, all of which have a maximum interpolation level equal to $L$.
As can easily be observed, isotropic sets are monotone.
The growth complexity of isotropic sparse grids is $\mathcal{O}\left(m \left(L\right) \left(\log m \left(L\right)\right)^{N-1}\right)$ \cite{bungartz_griebel_2004}.
While this complexity is much milder than the $\mathcal{O}\left(m\left(L\right)^N\right)$ of the full tensor-product collocation, isotropic collocation is obviously not free of the curse of dimensionality.
However, isotropic sparse grids can delay the curse of dimensionality up to a moderate number of \glspl{rv}.

\subsubsection{Adaptive anisotropic sparse grids}
\label{sec:adapt_sg} 
It is often the case that certain parameters or parameter combinations and interactions have a more significant impact on the \gls{qoi} than others.
This parameter anisotropy can be exploited to reduce the stochastic collocation's computational costs by using anisotropic sparse grids \cite{journals/siamnum/NobileTW08a}. 

Moreover, the anisotropic approximation can be constructed adaptively.
For that purpose, we will employ a greedy, dimension-adaptive algorithm, first presented in \cite{journals/computing/GerstnerG03} for quadrature purposes, and later used in \cite{journals/toms/KlimkeW05} for interpolation purposes.
In a \gls{uq} context and for the case of uniformly distributed input \glspl{rv}, similar approaches have been employed in \cite{journals/focm/ChkifaCS14, journals/jcphy/JakemanW15, journals/siamsc/NarayanJ14, SchiechePhD}.
The algorithm employed in this work is presented in Algorithm~\ref{algo:gensmolyak}.
An overview of the approach follows.

Let us assume that an approximation $\mathcal{I}_{\Lambda}\left[q\right]$ is readily available, where $\Lambda$ is a monotone multi-index set.
If not, we may initialize the dimension-adaptive Algorithm~\ref{algo:gensmolyak} with $\Lambda = \{\left(0,0,\dots,0\right)\}$.
All possible refinements of the multi-indices comprising $\Lambda$ form the refinement set
\begin{equation}
\label{eq:refinements}
 \mathcal{R}\left(\Lambda\right) = \left\{\boldsymbol{\ell}+\mathbf{e}_n, \forall \boldsymbol{\ell} \in \Lambda, \forall n=1,2,\dots,N \right\}.
\end{equation}
The admissible multi-indices, i.e. those that ensure the monotonicity property \eqref{eq:monotonicity} if added to $\Lambda$, form the admissible set
\begin{equation}
\label{eq:admissibility}
 \mathcal{A}\left(\Lambda\right) = \left\{\boldsymbol{\ell} \in \mathcal{R}\left(\Lambda\right) \: : \: \boldsymbol{\ell} \not\in \Lambda \: \text{and} \: \Lambda \cup \boldsymbol{\ell} \: \text{is monotone} \right\}.
\end{equation}
Each admissible multi-index $\boldsymbol{\ell} \in \mathcal{A}\left(\Lambda\right)$ defines a set of collocation points $\mathbf{y}_{\boldsymbol{\ell}}^{(\mathbf{i})} \in Z_{\boldsymbol{\ell}} \setminus Z_{\Lambda}$ which may be added to the available sparse grid $Z_\Lambda$.
The corresponding hierarchical surpluses are given by
\begin{equation}
  \label{eq:hs_multid}
  s_{\boldsymbol{\ell}}^{(\mathbf{i})} = q\left(\mathbf{y}_{\boldsymbol{\ell}}^{(\mathbf{i})}\right) - \mathcal{I}_{\Lambda}\left[q\right]\left(\mathbf{y}_{\boldsymbol{\ell}}^{(\mathbf{i})}\right), \:\:\: \mathbf{y}_{\boldsymbol{\ell}}^{(\mathbf{i})} \in Z_{\boldsymbol{\ell}} \setminus Z_{\Lambda}, \:\:\: \boldsymbol{\ell} \in \mathcal{A}\left(\Lambda\right).
\end{equation}
As proposed in \cite{journals/computing/GerstnerG03}, we use those hierarchical surpluses to compute the error indicators 
\begin{equation}
\label{eq:errorind}
 \eta_{\boldsymbol{\ell}} = \frac{1}{\# \left(Z_{\boldsymbol{\ell}} \setminus Z_{\Lambda}\right)} \sum_{\substack{ \mathbf{y}_{\boldsymbol{\ell}}^{(\mathbf{i})} \in Z_{\boldsymbol{\ell}} \setminus Z_{\Lambda} }} \left| s_{\boldsymbol{\ell}}^{(\mathbf{i})} \right|, \hspace{1em} \boldsymbol{\ell} \in \mathcal{A}\left(\Lambda\right).
\end{equation}
Other suggestions on error indicators can be found in \cite{journals/jcphy/JakemanW15, journals/siamsc/NarayanJ14, SchiechePhD}.
Finally, the multi-index set $\Lambda$ is extended with the admissible multi-index $\boldsymbol{\ell}^{*}$ corresponding to the maximum contribution, equivalently, the maximum $\eta_{\boldsymbol{\ell}}$.

The approximation $\mathcal{I}_{\Lambda^{*}}\left[q\right]$ can be constructed in a hierarchical way by adding the contributions of the new collocation points $\mathbf{y}_{\boldsymbol{\ell}^{*}}^{(\mathbf{i})} \in Z_{\boldsymbol{\ell}^{*}} \setminus Z_{\Lambda}$ to $\mathcal{I}_{\Lambda}\left[q\right]$, such that
\begin{align}
 \label{eq:hinterp_Nd}
 \mathcal{I}_{\Lambda^{*}}\left[q\right]\left(\mathbf{y}\right) &= \mathcal{I}_{\Lambda}\left[q\right]\left(\mathbf{y}\right) 
 + \sum_{\mathbf{i} : \mathbf{y}_{\boldsymbol{\ell}^{*}}^{(\mathbf{i})} \in Z_{\boldsymbol{\ell}^{*}} \setminus Z_{\Lambda}} \left(  q\left(\mathbf{y}_{\boldsymbol{\ell}^{*}}^{(\mathbf{i})}\right) - \mathcal{I}_{\Lambda}\left[q\right]\left(\mathbf{y}_{\boldsymbol{\ell}^{*}}^{(\mathbf{i})}\right) \right) L^{(\mathbf{i})}_{\boldsymbol{\ell}^{*}}\left(\mathbf{y}\right) \\
 &= \mathcal{I}_{\Lambda}\left[q\right]\left(\mathbf{y}\right) 
 + \sum_{\mathbf{i} : \mathbf{y}_{\boldsymbol{\ell}^{*}}^{(\mathbf{i})} \in Z_{\boldsymbol{\ell}^{*}} \setminus Z_{\Lambda}}   s_{\boldsymbol{\ell}^{*}}^{(\mathbf{i})} L^{(\mathbf{i})}_{\boldsymbol{\ell}^{*}}\left(\mathbf{y}\right) \nonumber.
\end{align}

The algorithm terminates either when a pre-defined simulation budget $B$, i.e. number of model evaluations, is reached, or when the total contribution of set $\mathcal{A}\left(\Lambda\right)$ is below a specified tolerance $\epsilon$, similarly to \cite{journals/computing/GerstnerG03, journals/jcphy/JakemanW15, journals/siamsc/NarayanJ14, journals/toms/KlimkeW05, SchiechePhD}.
The termination conditions can be formally formulated as
\begin{subequations}
\begin{align}
 \#  Z_{\Lambda} + \# Z_{\mathcal{A}\left(\Lambda\right)}  &\geq B, \\
 \sum_{\boldsymbol{\ell} \in \mathcal{A}\left(\Lambda\right)} \eta_{\boldsymbol{\ell}} &\leq \epsilon.
\end{align}
\end{subequations}
Since the hierarchical surpluses and collocation points for the admissible multi-indices have already been computed, the final approximation after the algorithm's termination is constructed with all multi-indices in the set $\Lambda \cup \mathcal{A}\left(\Lambda\right)$.

\begin{algorithm}[t]
 \SetAlgoLined
  \KwData{\gls{qoi} $q$, number of \glspl{rv} $N$, monotone multi-index set $\Lambda$, tolerance $\epsilon$, budget $B$}
  \KwResult{sparse approximation $\mathcal{I}_{\Lambda}\left[q\right]$}
  \Repeat{termination$\left[\mathcal{A}\left(\Lambda\right), \epsilon, B\right]$}{
  compute $\mathcal{A}\left(\Lambda\right)$ as in \eqref{eq:admissibility} \\
  compute $\eta_{\boldsymbol{\ell}}$, $\forall \boldsymbol{\ell} \in \mathcal{A}\left(\Lambda\right)$, as in \eqref{eq:errorind} \\
  find $\boldsymbol{\ell}^{*} = \text{arg}\max_{\boldsymbol{\ell} \in \mathcal{A}\left(\Lambda\right)} \eta_{\boldsymbol{\ell}}$ \\ 
  compute $\mathcal{I}_{\Lambda^{*}}\left[q\right]$ for $\Lambda^{*} = \Lambda \cup \boldsymbol{\ell}^{*}$, as in \eqref{eq:hinterp_Nd}\\
  set $\Lambda = \Lambda^{*}$
  } 
  set $\Lambda = \Lambda \cup \mathcal{A}\left(\Lambda\right)$
  \caption{Dimension-adaptive collocation}
  \label{algo:gensmolyak}
\end{algorithm}

\subsection{Post-Processing and Quadrature}
\label{sec:quad_postproc}
The approximation given by the stochastic collocation method can be used as an inexpensive surrogate model for sampling-based estimations of statistical measures, as proposed in Section~\ref{sec:setting}.
For example, assuming that a collocation based approximation $\mathcal{I}\left[q\right]$ is available, a statistical measure $\mathbb{E}\left[\phi\left(q\right)\right]$, as given in \eqref{eq:uq_task}, can be estimated in a \gls{mc} fashion as
\begin{equation}
 \mathbb{E}\left[\phi\left(q\right)\right] \approx M^{-1} \sum_{m=1}^M \phi\left(\mathcal{I}\left[q\right]\left(\mathbf{y}^{(m)}\right)\right),
\end{equation} 
where $M$ denotes the number of sampling points, respectively, model evaluations.
While the slow convergence of the \gls{mc} method remains, the computationally inexpensive polynomial surrogate model, $\mathcal{I}\left[q\right]$, allows us to draw large number of sample points $M$, thus significantly reducing the costs compared to sampling the original model. 

As a typical case where such a surrogate-based sampling approach would be useful, we consider a variance-based sensitivity analysis, commonly known as the Sobol method \cite{sobol2001}, based on the sampling-based algorithm suggested in \cite{saltelli2002}. 
Given $M$ randomly generated input realizations, the algorithm requires $\left(2N+2\right)M$ model evaluations to compute the sensitivity metrics, i.e. the Sobol indices.
For sufficient accuracy regarding the sensitivity metrics, $M$ is usually in the range of hundreds to thousands.
Further considering a large number of input \glspl{rv}, $N$, as well as a computationally expensive model, the overall cost of the analysis can become very high.
However, the task becomes feasible, if a sufficiently accurate surrogate model is used instead.

The aforementioned Sobol indices can be computed by directly post-processing an approximation with orthogonal terms, see e.g. \cite{sudret2008} for the case of generalized polynomial chaos approximations.
This non-sampling approach can still be used in combination with the hierarchical interpolation scheme presented in this work by transforming  \eqref{eq:hinterp_Nd} into an equivalent polynomial chaos expansion. 
This transformation is possible with a change of polynomial basis, i.e. by transforming the hierarchical polynomial basis into a basis of suitable Wiener-Askey polynomials \cite{journals/siamsc/XiuK02}, e.g. Legendre polynomials in the case of uniform input distributions.
An efficient method for this basis change is proposed in \cite{buzzard2013}.
A similar approach is described in \cite[Appendix A]{porta2014}.

Sampling is also not necessary for the estimation of statistical moments, which can be directly derived out of the approximation terms.
In this case, no change of basis is needed.
Considering first a univariate approximation of level $\ell$ as in \eqref{eq:hinterp_1d}, thus employing $m\left(\ell\right)$ collocation points, we apply the expectation operator such that the expected (mean) value of the \gls{qoi} can be estimated as
\begin{align}
\label{eq:exp_val_approx_1d}
 \mathbb{E}\left[q\right] = \int_{\Xi} q(y) \rho(y) \mathrm{d}y &\approx \int_{\Xi} \mathcal{I}\left[q\right](y) \rho(y) \mathrm{d}y = \int_{\Xi} \left(\sum_{i=1}^{m\left(\ell\right)} s_{\ell}^{(i)} l_{\ell}^{(i)}(y)\right) \rho(y) \mathrm{d}y \nonumber \\
 &= \sum_{i=1}^{m\left(\ell\right)} s_{\ell}^{(i)} \int_{\Xi} l_{\ell}^{(i)}(y) \rho(y) \mathrm{d}y 
 = \sum_{i=1}^{m\left(\ell\right)} s_{\ell}^{(i)} \mathbb{E}\left[ l_{\ell}^{(i)} \right].
\end{align}
Similar schemes can be used for the estimation of higher order moments, using approximations for the quantities $q^p$, where $p$ denotes the moment order \cite{SchiechePhD}.
We observe that \eqref{eq:exp_val_approx_1d} is similar to a univariate quadrature rule, where the function's evaluations on the quadrature nodes are incorporated in the coefficients $s_{\ell}^{(i)}$ and the quadrature weights coincide with $\mathbb{E}\left[l_{\ell}^{(i)}\right]$.
Thus, considering a continuous functional $\phi(q): \Xi \rightarrow \mathbb{R}$, we introduce a univariate quadrature rule, such that
\begin{equation}
 \label{eq:quad_1d} 
 \mathbb{E}\left[\phi(q)\right] \approx \mathcal{Q}_{\ell} \left[\phi(q)\right] = \sum_{i=1}^{m\left(\ell\right)} w_{\ell}^{(i)} \phi\left(q\left(y_{\ell}^{(i)}\right)\right),
\end{equation}
where the quadrature weights $w_{\ell}^{(i)}$ are given by
\begin{equation}
\label{eq:weight_1d}
 w_{\ell}^{(i)} = \int_{\Xi} l_{\ell}^{(i)}\left(y\right) \rho\left(y\right)\mathrm{d}y.
\end{equation}

Moving to the multivariate case, tensor-product quadrature rules can be constructed similarly to \eqref{eq:interp_tp}, such that
\begin{align}
 \label{eq:quad_tp}
 \mathcal{Q}_{\boldsymbol{\ell}} \left[\phi(q)\right] 
 &= \left(\mathcal{Q}_{1,\ell_1} \otimes \mathcal{Q}_{2,\ell_2} \otimes \cdots \otimes \mathcal{Q}_{N,\ell_N}\right) \left[\phi(q)\right] \nonumber \\ 
 &= \sum_{i_1=1}^{m_1\left(\ell_1\right)}\sum_{i_2=1}^{m_2\left(\ell_2\right)} \cdots \sum_{i_N=1}^{m_N\left(\ell_N\right)} \phi\left(q\left(y_{1,\ell_1}^{(i_1)}, y_{2,\ell_2}^{(i_2)}, \dots, y_{N,\ell_N}^{(i_N)}\right)\right)
 \prod_{n=1}^N w_{n, \ell_n}^{(i_n)} \\ \nonumber
 &= \sum_{\mathbf{i} : \mathbf{y}_{\boldsymbol{\ell}}^{(\mathbf{i})} \in Z_{\boldsymbol{\ell}}} \phi\left(q\left(\mathbf{y}_{\boldsymbol{\ell}}^{(\mathbf{i})}\right)\right) w^{(\mathbf{i})}_{\boldsymbol{\ell}}, \nonumber
\end{align}
where the multivariate weights $w^{(\mathbf{i})}_{\boldsymbol{\ell}}$ are given as products of the univariate ones, i.e.
\begin{equation}
\label{eq:weight_Nd}
  w^{(\mathbf{i})}_{\boldsymbol{\ell}} = \prod_{n=1}^N w_{n, \ell_n}^{(i_n)}.
\end{equation}
Then, assuming a readily available multivariate approximation based on a multi-index set with cardinality $\#\Lambda = K$, given as  
\begin{equation}
 \label{eq:kterm}
 q(\mathbf{y}) \approx \sum_{\boldsymbol{\ell} \in \Lambda} \Delta_{\boldsymbol{\ell}}\left[q\right]\left(\mathbf{y}\right) = \sum_{k=1}^K s_k L_k(\mathbf{y}),
\end{equation}
as well as an 1-1 relation between the global index $k$ and all combinations of the multi-indices $\boldsymbol{\ell}, \mathbf{i},$ corresponding to the collocation points $\mathbf{y}_{\boldsymbol{\ell}}^{(\mathbf{i})} \in Z_\Lambda$, the expected value of the \gls{qoi} can be estimated as
\begin{align}
\label{eq:exp_val_approx}
 \mathbb{E}\left[q\right] \approx \int_{\Xi} \left(\sum_{k=1}^K s_k L_k(\mathbf{y})\right) \rho(\mathbf{y})\mathrm{d}\mathbf{y}  
 = \sum_{k=1}^K s_k \int_{\Xi}  L_k(\mathbf{y}) \rho(\mathbf{y})\mathrm{d}\mathbf{y}
 = \sum_{k=1}^K s_k \mathbb{E}\left[L_k\right],
\end{align}
where the multivariate Lagrange polynomials $L_k$ are given as products of univariate ones, as in \eqref{eq:lagrangeNd}. 
Taking also into consideration \eqref{eq:weight_1d} and \eqref{eq:weight_Nd}, it holds that
\begin{equation}
 \mathbb{E}\left[L_k\right] = \mathbb{E}\left[L_{\boldsymbol{\ell}}^{(\mathbf{i})}\right] = \prod_{n=1}^N \mathbb{E}\left[l_{n, \ell_n}^{(i_n)}\right] = \prod_{n=1}^N w_{n, \ell_n}^{(i_n)} = w_{\boldsymbol{\ell}}^{(\mathbf{i})}.
\end{equation}
Therefore, \eqref{eq:exp_val_approx} is similar to a $K$-term multivariate quadrature rule, where the function's evaluations on the quadrature nodes are incorporated in the coefficients $s_k$ and the quadrature weights are given as products of univariate weights.

\subsection{Collocation Point Choices}
\label{sec:colloc_points}
As already pointed out, sparse grids, especially adaptive  ones, are based on nested univariate collocation grids.
Therefore, nestedness is a key requirement for the employed collocation points.
Moreover, the Lebesgue constant associated with the collocation points must remain bounded such that the interpolation yields accurate results \cite{journals/focm/ChkifaCS14}. 
Finally, the selected points must form accurate quadrature rules, to be used for the computation of statistical measures.
In this work we focus on two families of collocation points which satisfy all three requirements, namely the Clenshaw-Curtis and Leja nodes, described in the following.

\subsubsection{Clenshaw-Curtis collocation}
\label{sec:cc}

The first option is to use as univariate collocation points the nodes of the Clenshaw-Curtis quadrature rule. 
The rule has been proposed in \cite{clenshaw1960} for the integral approximation  
\begin{equation}
  \label{eq:integral_approx}
 \int_{-1}^1 q(y)\mathrm{d}y \approx \sum_{i=1}^n w^{(i)} q\left(y^{(i)}\right),
\end{equation}
where $y^{(i)}$ are the quadrature nodes and $w^{(i)}$ the quadrature weights.
The standard Clenshaw-Curtis nodes are extrema of Chebyshev polynomials $T_k(y)$ in the interval $\left[-1,1\right]$, plus the  boundary points of the interval \cite{trefethen2008}. 
The weights are typically computed by sums of trigonometric functions \cite{sommariva2013fast}.
Nested Clenshaw-Curtis nodes are obtained with the level-to-nodes function $m(\ell) = 2^{\ell} + 1$, with $m(0) = 1$, such that $Z_{\ell-1} \subset Z_{\ell}$, with $\#\left(Z_\ell \setminus Z_{\ell-1}\right) = 2^\ell$.

For integrations over general bounded domains $\left[a, b\right]$, the quadrature nodes and weights can be easily derived by simply scaling the nodes and weights in $\left[-1,1\right]$.
\begin{comment}
, such that
\begin{align}
 y'^{(i)} = \frac{b-a}{2}  y^{(i)} + \frac{a+b}{2}, \hspace{2em}
 w'^{(i)} = \frac{b-a}{2}  w^{(i)},
\end{align}
where $y'^{(i)}, w'^{(i)}$ refer to the interval $\left[a, b\right]$.
\end{comment}
It is therefore straightforward to extend the quadrature to integrals $\int_{a}^b q(y) \rho(y)\mathrm{d}y$
with a constant weight function $\rho(y)$, e.g. a uniform \gls{pdf} $\rho(y) = 1/\left(b-a\right)$ with support in $\left[a,b\right]$.
\begin{comment}
\begin{align}
 \int_{a}^b q(y) \rho(y)\mathrm{d}y 
 = \int_{a}^b q(y) \frac{1}{b-a}\mathrm{d}y 
 &\approx \frac{1}{b-a} \sum_{i=1}^n w'^{(i)} q\left(y'^{(i)}\right) \nonumber \\
 &= \frac{1}{b-a} \sum_{i=1}^n \frac{b-a}{2}w^{(i)} q\left(y'^{(i)}\right) 
 = \frac{1}{2} \sum_{i=1}^n  w^{(i)} q\left(y'^{(i)}\right).
\end{align}
\end{comment}

In the case of a non-uniform \gls{pdf} $\rho(y)$, or, generally, a non-constant weight function, the quadrature weights must be recomputed.
As already said, the nodes correspond to extrema of Chebyshev polynomials, and are therefore independent of the weight function.
A numerically efficient construction of non-uniform Clenshaw-Curtis weights has been given in \cite{sommariva2013fast}. 
The proposed approach is based on the discrete sine/cosine transform and is adopted in this work. 
To be precise, the $i$-th Clenshaw-Curtis weight is given by 
\begin{equation}
w^{(i)} = \frac{1}{2(i-1)} \left(2 \sum_{k=0}^{i-1} (-1)^k\gamma_k  + \gamma_0  + (-1)^i \gamma_{i+1}  \right), 
\end{equation}
where $\gamma_k = \int_{-1}^1 T_k(y) \rho(y) \ \mathrm{d}y$ represent moments of the Chebyshev polynomial $T_k$, to be precomputed.

\subsubsection{Leja collocation}
\label{sec:leja}

The second option is to base the collocation on Leja sequences.
The classic, unweighted Leja sequence is defined as a sequence of points $\left(y^{(i)}\right)_{i \geq 0}$, where $y^{(i)} \in \left[-1,1\right]$, $\forall i \geq 0$, such that
\begin{equation}
 \label{eq:leja_unweighted}
 y^{(i)} = \argmax_{y \in \left[-1,1\right]} \prod_{k=0}^{i-1} \left|y-y^{(k)}\right|,
\end{equation}
where the initial point $y^{(0)}$ can be chosen arbitrarily in $\left[-1,1\right]$ \cite{leja1957}.
Given a weight function $\rho(y)$ with support in a bounded interval $\left[a, b\right]$, weighted Leja sequences \cite{journals/siamsc/NarayanJ14} can be constructed as
\begin{equation}
 \label{eq:leja_weighted}
 y^{(i)} = \argmax_{y \in \left[a,b\right]} \sqrt{\rho(y)}\prod_{k=0}^{i-1} \left|y-y^{(k)}\right|.
\end{equation}
For weight functions corresponding to \glspl{pdf} of uniform distributions, we can still use the unweighted Leja rule \eqref{eq:leja_unweighted} with a simple scaling.
Although we do not consider unbounded domains in this work, we note that results concerning the Leja sequences in unbounded domains can be found in \cite{jantsch2018, journals/siamsc/NarayanJ14}.
Leja nodes are nested by definition, therefore, any level-to-nodes function satisfies the nestedness constraint.
In the context of this work we opt for $m(\ell) = \ell+1$, such that $\#\left(Z_\ell \setminus Z_{\ell-1}\right) = 1$.
Moreover, the weighted Leja nodes are tailored to the given weight function $\rho(y)$.
Finally, by integrating an interpolant constructed with a given Leja sequence, Leja-based quadrature rules can be constructed \cite{journals/siamsc/NarayanJ14}.
In the case of weighted Leja sequences, the corresponding quadrature weights are also tailored to the given weight function.

\section{Numerical Experiments}
\label{sec:numexp}
The aim of the following experiments is to assess the performance of Leja and Clenshaw-Curtis nodes in terms of interpolation and quadrature, in the context of the stochastic collocation method and for both uniform and bounded, non-uniform input densities.

In interpolation studies, the accuracy of the surrogate models is measured with three cross-validation error metrics.
Using a cross-validation set with $M$ random realizations of the input parameters, we compute the maximum absolute, mean absolute, and \gls{rms} errors
\begin{align}
 \epsilon_{\text{cv}, \text{max}} &= \max_{m=1,2,\dots,M} \left| \widetilde{q}\left(\mathbf{y}^{(m)}\right) - q\left(\mathbf{y}^{(m)}\right)\right|, \label{eq:cverr_max} \\
 \epsilon_{\text{cv}, \text{mean}} &= \frac{1}{M} \sum_{m=1}^M \left|\widetilde{q}\left(\mathbf{y}^{(m)}\right) - q\left(\mathbf{y}^{(m)}\right)\right|, \label{eq:cverr_mean} \\
 \epsilon_{\text{cv}, \text{RMS}} &= \sqrt{\frac{1}{M} \sum_{m=1}^M \left(\widetilde{q}\left(\mathbf{y}^{(m)}\right) - q\left(\mathbf{y}^{(m)}\right)\right)^2}, \label{eq:cverr_rms}
\end{align}
where $\widetilde{q}$ and $q$ denote the surrogate and the original model, respectively.
Both the $\epsilon_{\text{cv}, \text{mean}}$ and $\epsilon_{\text{cv}, \text{RMS}}$ metrics express the average error of the surrogate model.
From an interpretation point of view, $\epsilon_{\text{cv}, \text{mean}}$ is the most suitable measure of average performance.
However, the use of $\epsilon_{\text{cv}, \text{RMS}}$ has the benefit of penalizing large errors, thus being more sensitive to outliers, e.g. parameter realizations corresponding to the tail of the \gls{pdf}, or other regions in the parameter space where the approximation is not sufficiently accurate.
The error $\epsilon_{\text{cv}, \text{max}}$ can be seen as a measure for worst-case performance, being the most sensitive metric with respect to outliers.
In quadrature studies, the collocation-based estimates regarding the statistical moments of the considered \glspl{qoi} and the corresponding reference values are used to compute absolute or relative errors
\begin{align}
  \epsilon_{\text{abs}} &= \left| \mathbb{E}\left[\phi(q)\right]_{\text{ref}} - \mathbb{E}\left[\phi(q)\right] \right|, \label{eq:abserr}\\
  \epsilon_{\text{rel}} &= \left| \frac{\mathbb{E}\left[\phi(q)\right]_{\text{ref}} - \mathbb{E}\left[\phi(q)\right]}{\mathbb{E}\left[\phi(q)\right]_{\text{ref}}} \right| \label{eq:relerr}.
\end{align}

The interest in using quadrature methods based on Leja points is their straightforward construction for non-uniform densities, as presented in Section~\ref{sec:leja}. 
Although Clenshaw Curtis rules for arbitrary weights have been proposed recently \cite{sommariva2013fast}, their use in the \gls{uq} context is scarce. 
Hence, there is a strong interest in numerically comparing their performances.
Similarly, in the context of surrogate modeling, a comparison of approximation accuracies between surrogate models based on different node families is needed.

The univariate quadrature nodes and weights are computed with the Python package Chaospy \cite{journals/jocs/FeinbergL15}.
For non-uniform inputs, the Clenshaw-Curtis quadrature weights are adapted via the procedure suggested in \cite{sommariva2013fast}, using a self-developed implementation.
Univariate interpolations are based on the barycentric implementation \cite{berrut2004barycentric} provided by SciPy. 
In the multivariate case, we employ dimension-adaptive schemes, based on Algorithm~\ref{algo:gensmolyak}.
The dimension-adaptive Clenshaw-Curtis collocation employs the Sparse Grids MATLAB Kit \cite{csqi}. 
The software does not support non-uniform bounded distributions, and is therefore complemented by self-developed implementations when non-uniform inputs are considered.
An in-house, Python-based software \cite{loukrezis2018} is used for the dimension-adaptive Leja collocation.

\subsection{Dielectric Slab Waveguide} 
\label{sec:waveguide}

\begin{figure}[t!]
\centering
\includegraphics[width=0.8\textwidth]{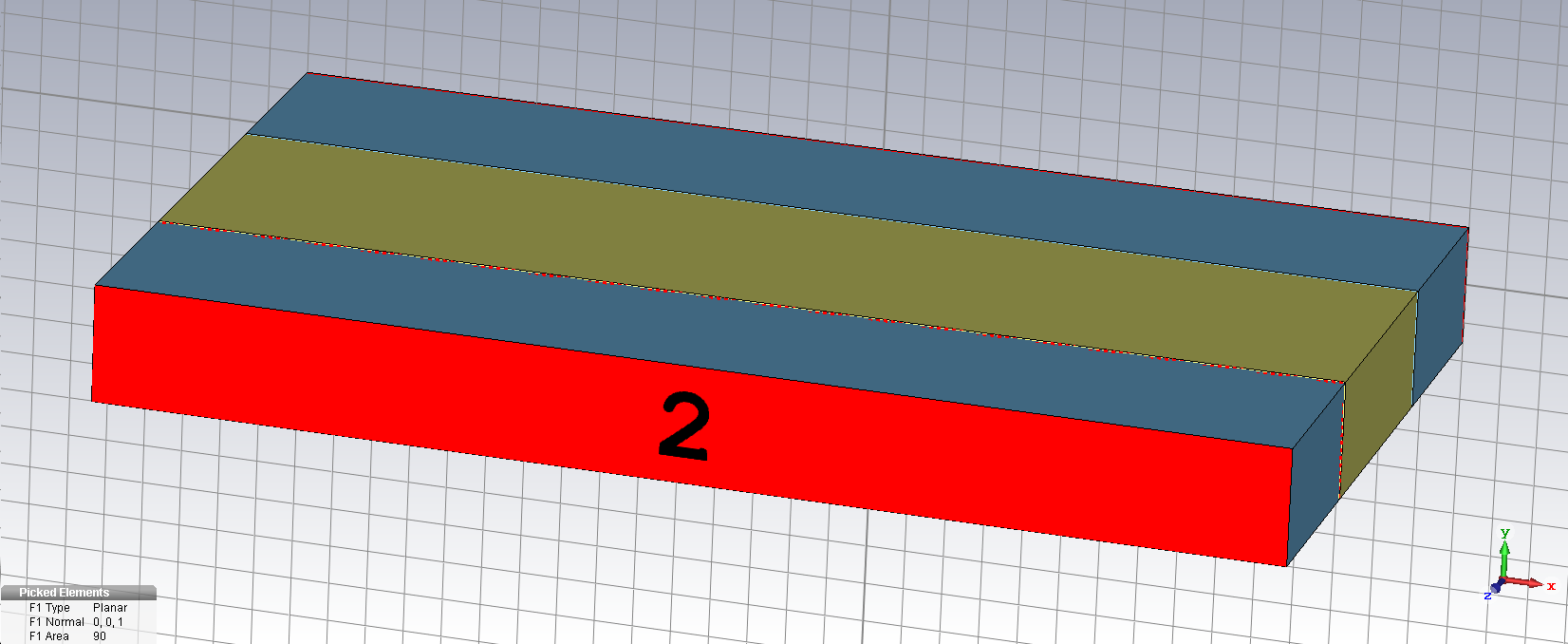}
\caption{3D waveguide model with dielectric filling. The yellow area denotes the dielectric filling and the blue area the vacuum. 
The red planes denote the waveguide ports.}
\label{fig:waveguide}
\end{figure}

We consider an academic example from the field of high-frequency electromagnetics. 
In particular, we consider a three-dimensional, rectangular, dielectric slab waveguide, as the one illustrated in Figure~\ref{fig:waveguide}.
The waveguide is filled with vacuum (blue area) and has a dielectric material with permittivity $\varepsilon = \varepsilon_0 \varepsilon_{\mathrm{r}}$ and permeability $\mu = \mu_0 \mu_{\mathrm{r}}$ in its middle (yellow area).
For both material properties, the subscript ``0'' refers to the property in vacuum and the subscript ``r'' to the relative value in the dielectric material.
The waveguide's geometry is defined by its width $w$ along the $x$-axis, its height $h$ along the $y$-axis and the vacuum offset $d$ and dielectric filling length $l$ along the $z$-axis. 
The red planes denote the waveguide's input and output ports, respectively port 1 and 2.
With the exception of the ports, the walls of the waveguide are considered to be perfect electrical conductors (PEC).
Using the Cartesian coordinate system, the waveguide's computational domain is defined as
\begin{subequations}
\begin{align}
 \Omega &= \left[0, w\right] \times \left[0, h\right] \times \left[0, 2d+l\right], \\
\Gamma_{\text{PEC}} &= \{(x,y,z) \in \partial \Omega  \; : \; z \neq 0 \wedge z \neq 2d+l\}, \\
 \Gamma_{\text{in}} &= \{(x,y,z) \in \partial \Omega  \; : \; z = 0\}, \\
 \Gamma_{\text{out}} &= \{(x,y,z) \in \partial \Omega  \; : \; z = 2d+l\},
\end{align}
\end{subequations}
such that $\partial \Omega = \Gamma_{\text{PEC}} \cup \Gamma_{\text{in}} \cup \Gamma_{\text{out}}$.
Such structures provide us with simple models, typically used to study wave confinement mechanisms.

We assume that the structure is excited at port 1 by an incoming plane wave $\mathbf{U}^{\text{inc}}$ with a given angular frequency $\omega = 2 \pi f$, where $f$ is the frequency.
We further assume that the incoming field coincides with the fundamental transverse electric mode TE$_{10}$ and that higher order modes are quickly attenuated in the structure.
In that case, Maxwell's source problem for the electric field $\mathbf{E}$ reads
\begin{subequations}
\begin{align}
\text{curl} \left(\mu^{-1} \text{curl} \mathbf{E}\right) - \omega^2 \varepsilon \mathbf{E} &= 0, &&\mathrm{in} \ \Omega, \\
\mathbf{E} \times \mathbf{n} &= 0, &&\mathrm{on} \ \Gamma_{\text{PEC}},\\
\mathbf{n} \times \text{curl} \mathbf{E} + \gamma \mathbf{n} \times \left(\mathbf{n} \times \mathbf{E} \right) &= \mathbf{U}^{\text{inc}}, &&\mathrm{on}\ \Gamma_{\text{in}}, \\
\mathbf{n} \times \text{curl} \mathbf{E} + \gamma \mathbf{n} \times \left(\mathbf{n} \times \mathbf{E} \right) &= 0, &&\mathrm{on} \ \Gamma_{\text{out}},
\end{align}
\label{eq:source}%
\end{subequations}
where $\mathbf{n}$ is the outwards-pointing normal vector, $\gamma = j k^{\text{inc}}$, and $k^{\text{inc}}$ refers to the wavenumber of $\mathbf{U}^{\text{inc}}$.
See \cite{jin2015finite} for details.

Typical \glspl{qoi} for waveguide devices are the so-called scattering parameters, S-parameters for short. 
The S-parameters quantify the reflection and transmission of the incoming field at the ports of the waveguide.
For example, the $S_{11}$ parameter, also referred to as the reflection coefficient, quantifies the reflection at port 1 and is given by
\begin{equation}
\label{eq:s11}
 S_{11} = C^\text{inc} \int_{\Gamma_\text{in}} \mathbf{E} \cdot \mathbf{e}_{10} \ \mathrm{d}x,
\end{equation}
where $C^\text{inc}$ is a normalization constant and $\mathbf{e}_{10} = \mathbf{e}_y \sin\frac{\pi x}{w}$ \cite{jin2015finite}, with $\mathbf{e}_y = \left(0, 1, 0\right)$ being the unit vector in the Cartesian $y$-direction.

In the most general case, we may consider uncertainties with respect to geometry, material or source parameters. 
Then, the parametric counterpart of \eqref{eq:source} reads
\begin{subequations}
\begin{align}
\text{curl} \left(\mu\left(\mathbf{y}\right)^{-1} \text{curl} \mathbf{E}\right) - \omega^2 \varepsilon(\mathbf{y}) \mathbf{E} &= 0, &&\mathrm{in} \ \Omega(\mathbf{y}), \\
\mathbf{E} \times \mathbf{n} &= 0, &&\mathrm{on} \ \Gamma_{\text{PEC}}(\mathbf{y}),\\
\mathbf{n} \times \text{curl} \mathbf{E} + \gamma \mathbf{n} \times \left(\mathbf{n} \times \mathbf{E} \right) &= \mathbf{U}^{\text{inc}}\left(\mathbf{y}\right), &&\mathrm{on}\ \Gamma_{\text{in}}(\mathbf{y}), \\
\mathbf{n} \times \text{curl} \mathbf{E} + \gamma \mathbf{n} \times \left(\mathbf{n} \times \mathbf{E} \right) &= 0, &&\mathrm{on} \ \Gamma_{\text{out}}(\mathbf{y}).
\end{align}
\label{eq:source_parametric}%
\end{subequations}
The solution of \eqref{eq:source_parametric} is also parameter-dependent, i.e. $\mathbf{E} = \mathbf{E}(\mathbf{y})$. 
Accordingly, the parametric $S_{11}$ parameter is given by
\begin{equation}
\label{eq:s11_param}
 S_{11}(\mathbf{y}) = C^\text{inc} \int_{\Gamma_\text{in}(\mathbf{y})} \mathbf{E}(\mathbf{y}) \cdot \mathbf{e}_{10} \ \mathrm{d}x.
\end{equation}
A common approach in the context of \gls{uq} with random geometries is to pull back the parametric equations to a fixed reference domain.
This approach ensures the tensor-product structure of the solution space.
However, since in this example we do not approximate the solution itself, but only a scalar \gls{qoi}, this transformation is not required.

In this particular example, the mathematical model is given by \eqref{eq:source}, where the frequency $f$ is now fixed at $6$ GHz.
The parametric model is given by \eqref{eq:source_parametric}, with $\mathbf{y} = \left(w,h,l,d,\varepsilon_{\text{r}}, \mu_{\text{r}}\right)$.
The parameter values for the nominal configuration of the waveguide are presented in Table~\ref{tab:wg_nom}.
The \gls{qoi} is chosen to be the magnitude of the waveguide's reflection coefficient at port 1, i.e. $q(\mathbf{y}) = \left|S_{11}\right|(\mathbf{y})$.
For this simple model, a semi-analytical solution for $S_{11}$ exists and is used so that we may avoid the consideration of discretization errors.

In the following, we will assume that the waveguide parameters are \glspl{rv} following either uniform or beta distributions.
In the uniform case, the distributions' lower and upper bounds for each parameter are given in Table~\ref{tab:wg_nom}.
We denote those bounds with $a_n$ and $b_n$, such that $Y_n \sim \mathcal{U}\left(a_n, b_n\right)$.
In the beta case, i.e. for $Y_n \sim \mathcal{B}\left(\alpha_n, \beta_n, a_n, b_n\right)$, the bounds, $a_n$ and $b_n$, coincide with the uniform ones.
The values of the shape parameters are $\alpha=3$ and $\beta=6$, resulting in a non-symmetric, positively skewed, i.e. right-tailed, distribution.
The shape parameters are the same for all \glspl{rv}, i.e. $\alpha_n = \alpha = 3$ and $\beta_n = \beta = 6$, for all $n=1,2,\dots,6$.

\begin{table}[b!]
\caption{Nominal parameter values and maximum deviations for the dielectric slab waveguide.}
\centering\begin{tabular}{|c|c|c|c|c|c|}
\hline
\textbf{Parameter} & \textbf{Symbol} & \textbf{Nominal Value} & \textbf{Lower Bound} & \textbf{Upper Bound} & \textbf{Units}  \\
\hline
  width & $w$ & $30$ & 27 & 33 & mm \\
\hline 
  height & $h$ & $3$ & 2.7 & 3.3 & mm \\
\hline
  filling length & $l$ & $7$ & 6.3 & 7.7 & mm \\
\hline
  vacuum offset & $d$ & $5$ & 4.5 & 5.5 & mm \\
\hline
  relative permittivity & $\varepsilon_{\mathrm{r}}$ & $2.0$ & 1.8 & 2.2 & -- \\
\hline
  relative permeability & $\mu_{\mathrm{r}}$ & $2.4$ & 2.16 & 2.64 &-- \\
\hline
\end{tabular}
\label{tab:wg_nom}
\end{table}

\subsubsection{Univariate quadrature results}
\label{sec:wgresults_quad_1d}

As a first test, we consider a single \gls{rv}, in particular, the waveguide's width $w$.
It is worth mentioning that results similar to the ones presented in the following have been obtained for the remaining waveguide parameters.
We compare quadrature errors in the expected value, variance and skewness, based on reference values computed with a Gauss rule with $30$ points.
The Clenshaw-Curtis rule is applied for quadrature levels $\ell=1,\dots, 4$, with $m\left( \ell \right) = 2^\ell + 1$ nodes per level.
The Leja rule employs quadrature levels $\ell=2, \dots, 16$, where $m\left( \ell \right) = i+ 1$.

Figure \ref{fig:1D-quadrature} depicts the absolute errors in the moments, for the case of a uniformly distributed parameter.
The Clenshaw-Curtis quadrature is consistently better for all three moments.
In the case of the expected value, the Clenshaw-Curtis rule has only a minor advantage over the Leja rule.
In the cases of the variance and the skewness, Leja quadrature remains competitive for accuracies up to $10^{-8}$ and $10^{-5}$, respectively.
For higher accuracies, the Clenshaw-Curtis rule is significantly better.
However, the obvious benefit of using the Leja rule is that, due to its more granular level-to-nodes function, it is able to offer increasing accuracies for numbers of nodes where the Clenshaw-Curtis is not nested, e.g. in between $10$ and $16$ quadrature nodes.

The results for the considered beta distribution are depicted in Figure \ref{fig:1D-quadrature_non-uniform}.
For the given choice of $\alpha=3, \beta=6$, and for all three moments, the weighted Leja quadrature is found to be slightly superior to the weighted Clenshaw Curtis rule, for the cases of $3$, $5$, and $9$ nodes.
The Clenshaw-Curtis rule again prevails when $17$ nodes are considered.
However, compared to the uniform case, the Leja rule remains competitive until much higher accuracies.
As before, the main advantage of the Leja rule is its nestedness property for an arbitrary number of nodes. 

\begin{figure}[t]
\centering     %%% not \center
     \subfigure[Expected value error.]{
          \pgfplotstableread[col sep = comma]{plot_data/quadrature/wg_uniform_quad_1D_Gauss.csv}\Gauss
	  \pgfplotstableread[col sep = comma]{plot_data/quadrature/wg_uniform_quad_1D_CC.csv}\CC
	  \pgfplotstableread[col sep = comma]{plot_data/quadrature/wg_uniform_quad_1D_Leja.csv}\Leja
       \begin{tikzpicture}
	\begin{semilogyaxis}[width=0.33\textwidth, xlabel={Model evaluations}, ylabel={$\epsilon_{\text{abs}}$},  legend pos=north east, font = \small, mark size = 1pt]
	  %\addplot[mark=square*, black, dashed] table [x index = {0}, y index = {1}] {\Gauss};
	  \addplot[mark=*, gray] table [x index = {0}, y index = {1}] {\CC};
	  \addplot[mark=*, black] table [x index = {0}, y index = {1}] {\Leja};
	  \legend{CC, L}
	\end{semilogyaxis}
      \end{tikzpicture}
      \label{fig:quad_1D_uniform_mean}
  }
  \hfill
  \subfigure[Variance error.]{
     \pgfplotstableread[col sep = comma]{plot_data/quadrature/wg_uniform_quad_1D_Gauss.csv}\Gauss
     \pgfplotstableread[col sep = comma]{plot_data/quadrature/wg_uniform_quad_1D_CC.csv}\CC
     \pgfplotstableread[col sep = comma]{plot_data/quadrature/wg_uniform_quad_1D_Leja.csv}\Leja
      \begin{tikzpicture}
	\begin{semilogyaxis}[width=0.33\textwidth, xlabel={Model evaluations}, legend pos=north east, font = \small, mark size = 1pt]
	  %\addplot table [x index = {0}, y index = {2}] {\Gauss};
	  \addplot[mark=*, gray] table [x index = {0}, y index = {2}] {\CC};
	  \addplot[mark=*, black] table [x index = {0}, y index = {2}] {\Leja};
	  \legend{CC, L}
	\end{semilogyaxis}
      \end{tikzpicture}
      \label{fig:quad_1D_uniform_var}
  }
  \hfill
    \subfigure[Skewness error.]{
     \pgfplotstableread[col sep = comma]{plot_data/quadrature/wg_uniform_quad_1D_Gauss.csv}\Gauss
     \pgfplotstableread[col sep = comma]{plot_data/quadrature/wg_uniform_quad_1D_CC.csv}\CC
     \pgfplotstableread[col sep = comma]{plot_data/quadrature/wg_uniform_quad_1D_Leja.csv}\Leja
      \begin{tikzpicture}
	\begin{semilogyaxis}[width=0.33\textwidth, xlabel={Model evaluations}, legend pos=north east, legend style={fill=none}, font = \small, mark size = 1pt]
	  %\addplot table [x index = {0}, y index = {3}] {\Gauss};
	  \addplot[mark=*, gray] table [x index = {0}, y index = {3}] {\CC};
	  \addplot[mark=*, black] table [x index = {0}, y index = {3}] {\Leja};
	  \legend{CC, L}
	\end{semilogyaxis}
      \end{tikzpicture}
      \label{fig:quad_1D_uniform_skew}
  }
\caption{Absolute moment errors for the dielectric slab waveguide with a single, uniform random input. The approximations are based on Clenshaw-Curtis and uniform (unweighted) Leja rules. The reference moment values are computed using a Gauss-Legendre rule with 30 nodes.}
\label{fig:1D-quadrature}
\end{figure}  

\begin{figure}[t]
\centering     %%% not \center
   \subfigure[Expected value error.]{
   \pgfplotstableread[col sep = comma]{plot_data/quadrature/wg_non-uniform_quad_1D_Gauss.csv}\Gauss
   \pgfplotstableread[col sep = comma]{plot_data/quadrature/wg_non-uniform_quad_1D_CC.csv}\CC
   \pgfplotstableread[col sep = comma]{plot_data/quadrature/wg_non-uniform_quad_1D_Leja.csv}\Leja
       \begin{tikzpicture}
	\begin{semilogyaxis}[width=0.33\textwidth, xlabel={Model evaluations}, ylabel=$\epsilon_{\text{abs}}$, legend style={fill=none}, legend pos=north east, font=\small, mark size=1pt]
	  %\addplot table [x index = {0}, y index = {1}] {\Gauss};
	  \addplot[mark=*, gray] table [x index = {0}, y index = {1}] {\CC};
	  \addplot[mark=*, black] table [x index = {0}, y index = {1}] {\Leja};
	  \legend{CC, L}
	\end{semilogyaxis}
      \end{tikzpicture}
      \label{fig:quad_1D_non-uniform_mean}
  }
  \hfill
  \subfigure[Variance error.]{
   \pgfplotstableread[col sep = comma]{plot_data/quadrature/wg_non-uniform_quad_1D_Gauss.csv}\Gauss
   \pgfplotstableread[col sep = comma]{plot_data/quadrature/wg_non-uniform_quad_1D_CC.csv}\CC
   \pgfplotstableread[col sep = comma]{plot_data/quadrature/wg_non-uniform_quad_1D_Leja.csv}\Leja
      \begin{tikzpicture}
	\begin{semilogyaxis}[width=0.33\textwidth, xlabel={Model evaluations}, legend style={fill=none}, legend pos=north east, font=\small, mark size=1pt]
	  %\addplot table [x index = {0}, y index = {2}] {\Gauss};
	  \addplot[mark=*, gray] table [x index = {0}, y index = {2}] {\CC};
	  \addplot[mark=*, black] table [x index = {0}, y index = {2}] {\Leja};
	  \legend{CC, L}
	\end{semilogyaxis}
      \end{tikzpicture}
      \label{fig:quad_1D_non-uniform_var}
  }
  \hfill
  \subfigure[Sknewness error.]{
   \pgfplotstableread[col sep = comma]{plot_data/quadrature/wg_non-uniform_quad_1D_Gauss.csv}\Gauss
   \pgfplotstableread[col sep = comma]{plot_data/quadrature/wg_non-uniform_quad_1D_CC.csv}\CC
   \pgfplotstableread[col sep = comma]{plot_data/quadrature/wg_non-uniform_quad_1D_Leja.csv}\Leja
      \begin{tikzpicture}
	\begin{semilogyaxis}[width=0.33\textwidth, xlabel={Model evaluations}, legend style={fill=none}, legend pos=north east, font=\small, mark size=1pt]
	  %\addplot table [x index = {0}, y index = {3}] {\Gauss};
	  \addplot[mark=*, gray] table [x index = {0}, y index = {3}] {\CC};
	  \addplot[mark=*, black] table [x index = {0}, y index = {3}] {\Leja};
	  \legend{CC, L}
	\end{semilogyaxis}
      \end{tikzpicture}
      \label{fig:quad_1D_non-uniform_skew}
  }
\caption{Absolute moment errors for the dielectric slab waveguide with a single, beta-distributed random input. The beta distribution's shape parameters are $\alpha=3$ and $\beta=6$. The approximations are based on a Clenshaw-Curtis rule with modified weights and a beta-weighted Leja rule. The reference moment values are computed using a Gauss-Jacobi rule with 30 nodes.}
\label{fig:1D-quadrature_non-uniform}
\end{figure}

\subsubsection{Univariate interpolation results}
\label{sec:wgresults_interp_1d}

Similarly to Section~\ref{sec:wgresults_quad_1d}, we investigate the performance of Leja and Clenshaw-Curtis, now in the interpolation context, i.e. with respect to approximation accuracy.
We consider again a single input \gls{rv}, in particular the waveguide's width $w$.

We first consider a uniformly distributed input \gls{rv} and construct two approximations, based on Clenshaw-Curtis and uniform (unweighted) Leja interpolation nodes, respectively.
We measure the approximation accuracy with cross-validation errors \eqref{eq:cverr_max}, \eqref{eq:cverr_mean}, and \eqref{eq:cverr_rms} using a validation set with $10^5$ parameter realizations drawn from the uniform \gls{pdf}.
The error-cost relation with respect to all three cross-validation errors is depicted in Figure~\ref{fig:1D-interpolation-uniform}.
The two rules show comparable performance, since their accuracy is almost indistinguishable in all error metrics, for the same number of model evaluations.
As in the quadrature case, the granularity of the Leja rule results in increasingly more accurate approximations for numbers of nodes for which the Clenshaw-Curtis rule is not nested.

Next we consider an input \gls{rv} which follows a positively skewed beta distribution ($\alpha=3$, $\beta=6$).
As already mentioned in Section~\ref{sec:cc}, the Clenshaw-Curtis nodes do not depend on the \gls{pdf} of the input parameter and remain unchanged in a given interval $\left[a,b\right]$.
On the contrary, weighted Leja nodes change according to the input \gls{pdf}, due to their definition in \eqref{eq:leja_weighted}.
Figure~\ref{fig:1D-interpolation-beta} shows the error-cost relation for approximations based on Clenshaw-Curtis and beta-weighted Leja rules.
A validation sample with $10^5$ random realizations drawn from the beta distribution is used to compute the cross-validation errors.

Contrary to the uniform case, we observe now that the weighted Leja rule is superior when the $\epsilon_{\text{cv}, \text{mean}}$ and $\epsilon_{\text{cv}, \text{RMS}}$ metrics are considered, but inferior in terms of the $\epsilon_{\text{cv}, \text{max}}$ metric.
As discussed in Section~\ref{sec:quad_postproc}, the errors $\epsilon_{\text{cv}, \text{mean}}$ and $\epsilon_{\text{cv}, \text{RMS}}$ quantify the expected performance of a surrogate model in terms of approximation accuracy, while $\epsilon_{\text{cv}, \text{max}}$ its worst-case performance.
Since the weighted Leja rule is tailored to the input beta distribution, its superior on-average performance is to be expected.
Its inferior $\epsilon_{\text{cv}, \text{max}}$ errors can possibly be explained by poor approximations regarding outliers, e.g. parameter realizations corresponding to the tails of the beta \gls{pdf}.

We note that the Clenshaw-Curtis-based results of Figure~\ref{fig:1D-interpolation-beta} are very similar to the ones obtained with a uniform (unweighted) Leja rule, always using the beta-generated validation sample for computing the cross-validation errors.
This result indicates that uniform-based approximations could be advantageous if the approximation of outliers or the worst-case performance of the surrogate model are of interest.
In this case, it might be computationally more efficient to construct an approximation for a uniform distribution and then sample it with realizations drawn from the true input distribution, instead of employing nodes suited specifically to the true input distribution.

\begin{figure}[t!]
\centering
  \subfigure[Maximum absolute error.]{
   \begin{tikzpicture} 
      \begin{semilogyaxis}[width=0.33\textwidth, xlabel=Model evaluations, ylabel=$\epsilon_{\text{cv}}$, legend style={fill=none}, legend pos=north east, font=\small, mark size=1pt]
	\addplot[mark=*, gray] table[x index=0, y index=1]{plot_data/interpolation/cverrz_1d/cverrz_cc_usample.txt};
	\addplot[mark=*, black] table[x index=0, y index=1]{plot_data/interpolation/cverrz_1d/cverrz_ul_usample.txt};
	\legend{CC, L}
      \end{semilogyaxis}
    \end{tikzpicture}
  }
  \hfill
  \subfigure[Mean absolute error.]{
   \begin{tikzpicture} 
      \begin{semilogyaxis}[width=0.33\textwidth, xlabel=Model evaluations, legend style={fill=none}, legend pos=north east, font=\small, mark size=1pt]
	\addplot[mark=*, gray] table[x index=0, y index=2]{plot_data/interpolation/cverrz_1d/cverrz_cc_usample.txt};
	\addplot[mark=*, black] table[x index=0, y index=2]{plot_data/interpolation/cverrz_1d/cverrz_ul_usample.txt};
	\legend{CC, L}
      \end{semilogyaxis}
    \end{tikzpicture}
  }
  \hfill
  \subfigure[Root-mean-square error.]{
   \begin{tikzpicture} 
      \begin{semilogyaxis}[width=0.33\textwidth, xlabel=Model evaluations, legend style={fill=none}, legend pos=north east, font=\small, mark size=1pt]
	\addplot[mark=*, gray] table[x index=0, y index=3]{plot_data/interpolation/cverrz_1d/cverrz_cc_usample.txt};
	\addplot[mark=*, black] table[x index=0, y index=3]{plot_data/interpolation/cverrz_1d/cverrz_ul_usample.txt};
	\legend{CC, L}
      \end{semilogyaxis}
    \end{tikzpicture}
  }
\caption{Cross-validation errors for the dielectric slab waveguide with a single, uniform random input. The approximations are based on Clenshaw-Curtis and uniform (unweighted) Leja rules. A validation set with $10^5$ parameter realizations is used.}
\label{fig:1D-interpolation-uniform}
\end{figure}

\begin{figure}[t!]
\centering
  \subfigure[Maximum absolute error.]{
   \begin{tikzpicture} 
      \begin{semilogyaxis}[width=0.33\textwidth, xlabel=Model evaluations, ylabel=$\epsilon_{\text{cv}}$, legend style={fill=none}, legend pos=north east, font=\small, mark size=1pt]
	\addplot[mark=*, gray] table[x index=0, y index=1]{plot_data/interpolation/cverrz_1d/cverrz_cc_b36sample.txt};
	\addplot[mark=*, black] table[x index=0, y index=1]{plot_data/interpolation/cverrz_1d/cverrz_bl_b36sample.txt};
	\legend{CC, L}
      \end{semilogyaxis}
    \end{tikzpicture}
  }
  \hfill
  \subfigure[Mean absolute error.]{
   \begin{tikzpicture} 
      \begin{semilogyaxis}[width=0.33\textwidth, xlabel=Model evaluations, legend style={fill=none}, legend pos=north east, font=\small, mark size=1pt]
	\addplot[mark=*, gray] table[x index=0, y index=2]{plot_data/interpolation/cverrz_1d/cverrz_cc_b36sample.txt};
	\addplot[mark=*, black] table[x index=0, y index=2]{plot_data/interpolation/cverrz_1d/cverrz_bl_b36sample.txt};
	\legend{CC, L}
      \end{semilogyaxis}
    \end{tikzpicture}
  }
  \hfill
  \subfigure[Root-mean-square error.]{
   \begin{tikzpicture} 
      \begin{semilogyaxis}[width=0.33\textwidth, xlabel=Model evaluations, legend style={fill=none}, legend pos=north east, font=\small, mark size=1pt]
	\addplot[mark=*, gray] table[x index=0, y index=3]{plot_data/interpolation/cverrz_1d/cverrz_cc_b36sample.txt};
	\addplot[mark=*, black] table[x index=0, y index=3]{plot_data/interpolation/cverrz_1d/cverrz_bl_b36sample.txt};
	\legend{CC, L}
      \end{semilogyaxis}
    \end{tikzpicture}
  }
\caption{Cross-validation errors for the dielectric slab waveguide with a single, beta-distributed random input. The approximations are based on Clenshaw-Curtis and beta-weighted Leja rules. A validation set with $10^5$ parameter realizations is used.}
\label{fig:1D-interpolation-beta}
\end{figure}

\subsubsection{Multivariate quadrature results}
\label{sec:wgresults_quad_Nd}

We now consider the multivariate case, which consists of all $6$ waveguide parameters, such that $\mathbf{Y} = \left(w,h,l,d,\varepsilon_{\mathrm{r}}, \mu_{\mathrm{r}}\right)$.
We apply the dimension-adaptive stochastic collocation method with Clenshaw-Curtis and Leja nodes, and for uniform and beta input distributions.
The reference moment values are obtained with tensor-product Gauss quadrature rules using $11$ nodes per parameter, thus resulting in $11^6$ model evaluations.

First, we assume that all parameters follow uniform distributions with the bounds shown in Table~\ref{tab:wg_nom}.
The results regarding the expected value and the variance of the \gls{qoi} are presented in Figure~\ref{fig:rel_errz_wg_uniform}.
As can be observed, the Clenshaw-Curtis-based adaptive scheme outperforms the Leja-based one for both moments.
The Leja rule can be seen as relatively competitive, however, the advantage of Clenshaw-Curtis is obvious, especially with respect to the variance.
This result coincides with similar observations from \cite{journals/siamsc/NarayanJ14}, where Leja nodes are found to be inferior to Clenshaw-Curtis nodes in sparse quadrature schemes.

Next, we consider the input \glspl{rv} to follow beta distributions with shape parameters $\alpha=3$, $\beta=6$, and bounded by the values given in Table~\ref{tab:wg_nom}. 
The relative error results are presented in Figure~\ref{fig:rel_errz_wg_beta36}. 
Contrary to the uniform case, the weighted Leja rule is significantly superior to the Clenshaw-Curtis rule with modified weights, for both moments.

\begin{figure}[t!]
\centering
  \subfigure[Relative errors for the expected value. \label{fig:rel_errz_mean_wg_uniform}]{
  \begin{tikzpicture} 
      \begin{semilogyaxis}[width=0.45\textwidth, xlabel=Model evaluations, ylabel=$\epsilon_{\text{rel}}$, legend pos=north east]
	\addplot[mark=None, gray, thick] table[x index=0, y index=1, col sep=comma]{plot_data/quadrature/quad_6d_adaptCC_uniform_relerrz.txt};
	\addplot[mark=None, black, thick] table[x index=0, y index=1]{plot_data/quadrature/quad_6d_dali_uniform_relerrz.txt};
	\legend{Clenshaw-Curtis, Leja}
      \end{semilogyaxis}
  \end{tikzpicture}
  }
  \subfigure[Relative errors for the variance. \label{fig:rel_errz_var_wg_uniform}]{
  \begin{tikzpicture} 
      \begin{semilogyaxis}[width=0.45\textwidth, xlabel=Model evaluations, legend pos=north east] 
	\addplot[mark=None, gray, thick] table[x index=0, y index=2, col sep= comma]{plot_data/quadrature/quad_6d_adaptCC_uniform_relerrz.txt};
	\addplot[mark=None, black, thick] table[x index=0, y index=2]{plot_data/quadrature/quad_6d_dali_uniform_relerrz.txt};
	\legend{Clenshaw-Curtis, Leja}
      \end{semilogyaxis}
  \end{tikzpicture}
  }
\caption{Relative moment errors for the dielectric slab waveguide with $6$ uniform random inputs. The approximations are based on Clenshaw-Curtis and uniform (unweighted) Leja rules. The reference moment values are computed using a tensor-product Gauss-Legendre rule with 11 nodes per parameter.}
\label{fig:rel_errz_wg_uniform}
\end{figure}

\begin{figure}[t!]
\centering
  \subfigure[Relative errors for the expected value. \label{fig:rel_errz_mean_wg_beta36}]{
  \begin{tikzpicture} 
      \begin{semilogyaxis}[width=0.45\textwidth, xlabel=Model evaluations, ylabel=$\epsilon_{\text{rel}}$, legend pos=north east]
	\addplot[mark=None, gray, thick] table[x index=0, y index=1, col sep=comma]{plot_data/quadrature/quad_6d_adaptCC_beta36_relerrz.txt};
	\addplot[mark=None, black, thick] table[x index=0, y index=1]{plot_data/quadrature/quad_6d_dali_beta36_relerrz.txt};
	\legend{Clenshaw-Curtis, Leja}
      \end{semilogyaxis}
  \end{tikzpicture}
  }
  \subfigure[Relative errors for the variance. \label{fig:rel_errz_var_wg_beta36}]{
  \begin{tikzpicture} 
      \begin{semilogyaxis}[width=0.45\textwidth, xlabel=Model evaluations, legend pos=north east] 
	\addplot[mark=None, gray, thick] table[x index=0, y index=2, col sep= comma]{plot_data/quadrature/quad_6d_adaptCC_beta36_relerrz.txt};
	\addplot[mark=None, black, thick] table[x index=0, y index=2]{plot_data/quadrature/quad_6d_dali_beta36_relerrz.txt};
	\legend{Clenshaw-Curtis, Leja}
      \end{semilogyaxis}
  \end{tikzpicture}
  }
\caption{Relative moment errors for the dielectric slab waveguide with $6$ beta-distributed random inputs. The shape parameters of all beta distributions are $\alpha=3$ and $\beta=6$. The approximations are based on a Clenshaw-Curtis rule with modified weights and a beta-weighted Leja rule. The reference moment values are computed using a tensor-product Gauss-Jacobi rule with 11 nodes per parameter.}
\label{fig:rel_errz_wg_beta36}
\end{figure}

\subsubsection{Multivariate interpolation results}
\label{sec:wgresults_interp_Nd}

Again, we compare the Clenshaw-Curtis and Leja rules in terms of interpolation accuracy, now considering all $6$ waveguide parameters.
As in the univariate case, we use the $\epsilon_{\text{cv}, \text{max}}$ error defined in \eqref{eq:cverr_max} to quantify the worst-case performance of both rules.
The average performance is measured with the $\epsilon_{\text{cv}, \text{RMS}}$ metric, defined in \eqref{eq:cverr_rms}.
We note that the results with respect to the $\epsilon_{\text{cv}, \text{mean}}$ metric are very similar to the ones of $\epsilon_{\text{cv}, \text{RMS}}$ and are therefore omitted. 

We first consider the case where all $6$ parameters follow uniform distributions with the bounds given in Table~\ref{tab:wg_nom}.
We apply the dimension-adaptive stochastic collocation method presented in Algorithm~\ref{algo:gensmolyak}, using Clenshaw-Curtis and uniform (unweighted) Leja rules.
The cross-validation errors are computed using a validation sample with $10^5$ parameter realizations, drawn from the uniform-based joint \gls{pdf}.
The corresponding results are presented in Figure~\ref{fig:MD-interpolation-uniform}.
As can be observed, the Leja rule outperforms the Clenshaw-Curtis rule for both error metrics, i.e. in terms of both average and worst-case performance.

Next we consider that all $6$ parameters follow positively skewed beta distributions with identical shape parameters ($\alpha=3$, $\beta=6$) and bounded by the lower and upper limits given in Table~\ref{tab:wg_nom}.
The dimension-adaptive scheme is now based on Clenshaw-Curtis and beta-weighted Leja rules and the cross-validation set is now drawn from the beta-based joint \gls{pdf}.
Again, $10^5$ parameter realizations are used to compute the cross-validation errors.
We also compute cross-validation errors using an unweighted Leja rule, in order to investigate whether uniform-based approximations could be advantageous in terms of worst-case performance, as was observed in the univariate case (see Section~\ref{sec:wgresults_interp_1d}).

Figure~\ref{fig:MD-interpolation-beta} shows the corresponding results.
As expected, the on-average performance of the beta-weighted Leja rule is significantly better than the other two options.
Contrary to the univariate case, the beta-weighted Leja rule outperforms both the Clenshaw-Curtis and the unweighted Leja rules also in terms of worst-case performance, as quantified by the $\epsilon_{\text{cv}, \text{max}}$ error.
Hence, in the multivariate case, no advantage is observed in using a uniform-based approximation instead of one dedicated to the joint input \gls{pdf}.

\begin{figure}[t!]
\centering
  \subfigure[Maximum absolute error.]{
   \begin{tikzpicture} 
      \begin{semilogyaxis}[width=0.45\textwidth, xlabel=Model evaluations, ylabel=$\epsilon_{\text{cv}}$, legend style={fill=none}, legend pos=north east]
	\addplot[mark=none, gray, thick] table[x index=0, y index=1]{plot_data/interpolation/results_6d/results_6d_cc_usample.txt};
	\addplot[mark=none, black, thick] table[x index=0, y index=1]{plot_data/interpolation/results_6d/results_6d_ul_usample.txt};
	\legend{Clenshaw-Curtis, Leja}
      \end{semilogyaxis}
    \end{tikzpicture}
  }
  \subfigure[Root-mean-square error.]{
   \begin{tikzpicture} 
      \begin{semilogyaxis}[width=0.45\textwidth, xlabel=Model evaluations, legend style={fill=none}, legend pos=north east]
	\addplot[mark=none, gray, thick] table[x index=0, y index=3]{plot_data/interpolation/results_6d/results_6d_cc_usample.txt};
	\addplot[mark=none, black, thick] table[x index=0, y index=3]{plot_data/interpolation/results_6d/results_6d_ul_usample.txt};
	\legend{Clenshaw-Curtis, Leja}
      \end{semilogyaxis}
    \end{tikzpicture}
  }
\caption{Cross-validation errors for the dielectric slab waveguide with 6 uniform random inputs. The approximations are based on Clenshaw-Curtis and uniform (unweighted) Leja rules. A validation set with $10^5$ parameter realizations is used.}
\label{fig:MD-interpolation-uniform}
\end{figure}

\begin{figure}[t!]
\centering
  \subfigure[Maximum absolute error.]{
   \begin{tikzpicture} 
      \begin{semilogyaxis}[width=0.45\textwidth, xlabel=Model evaluations, ylabel=$\epsilon_{\text{cv}}$, legend style={fill=none}, legend pos=north east]
	\addplot[mark=none, gray, thick] table[x index=0, y index=1]{plot_data/interpolation/results_6d/results_6d_cc_bsample.txt};
	\addplot[mark=none, black, thick] table[x index=0, y index=1]{plot_data/interpolation/results_6d/results_6d_bl_bsample.txt};
	\addplot[mark=none, black, thick, dashed] table[x index=0, y index=1]{plot_data/interpolation/results_6d/results_6d_ul_bsample.txt};
	\legend{Clenshaw-Curtis, weighted Leja, unweighted Leja}
      \end{semilogyaxis}
    \end{tikzpicture}
  }
  \subfigure[Root-mean-square error.]{
   \begin{tikzpicture} 
      \begin{semilogyaxis}[width=0.45\textwidth, xlabel=Model evaluations, legend style={fill=none}, legend pos=north east]
	\addplot[mark=none, gray, thick] table[x index=0, y index=3]{plot_data/interpolation/results_6d/results_6d_cc_bsample.txt};
	\addplot[mark=none, black, thick] table[x index=0, y index=3]{plot_data/interpolation/results_6d/results_6d_bl_bsample.txt};
	\addplot[mark=none, black, thick, dashed] table[x index=0, y index=1]{plot_data/interpolation/results_6d/results_6d_ul_bsample.txt};
	\legend{Clenshaw-Curtis, weighted Leja, unweighted Leja}
      \end{semilogyaxis}
    \end{tikzpicture}
  }
\caption{Cross-validation errors for the dielectric slab waveguide with 6 beta-distributed random inputs. The shape parameters of all beta distributions are $\alpha=3$ and $\beta=6$. The approximations are based on Clenshaw-Curtis, uniform (unweighted), and beta-weighted Leja rules. A validation set with $10^5$ parameter realizations is used.}
\label{fig:MD-interpolation-beta}
\end{figure}

\subsection{Stern-Gerlach Magnet}
\label{sec:sterngerlach}
\begin{figure}[t]
\centering     %%% not \center
\subfigure[3D magnet model.]{\includegraphics[width=0.45\textwidth, angle=0]{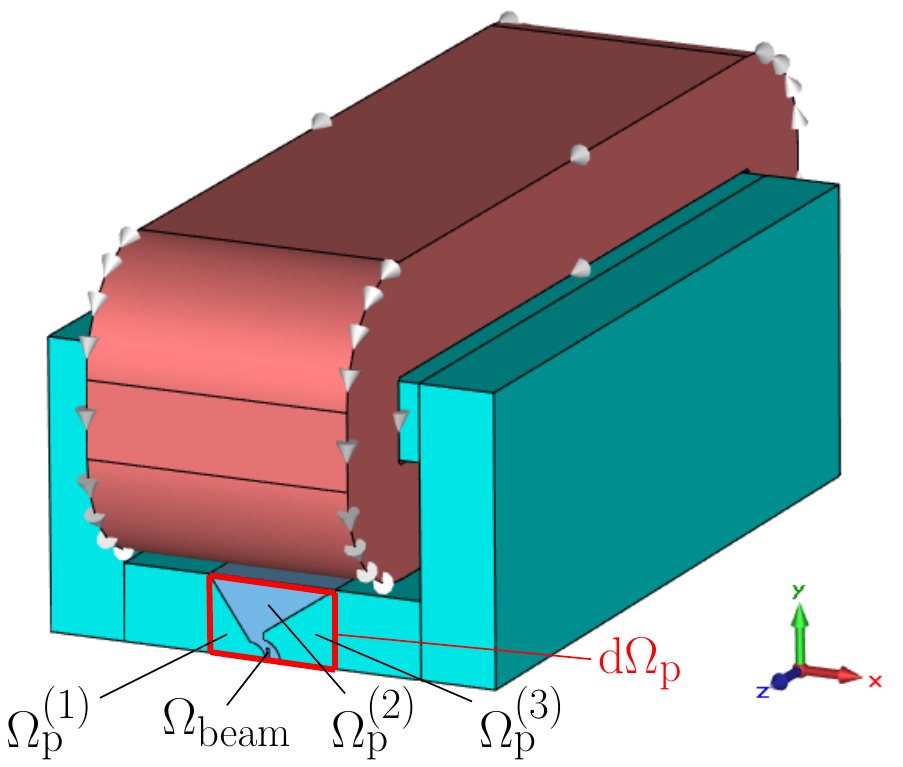} \label{fig:3dmodel}}
\subfigure[Zoom in magnet's pole region.]{
\begin{tikzpicture}[scale=0.4]
    \begin{axis}[
      width=1.2\columnwidth,
      height=0.8\columnwidth,
      cycle list name=color list,
      xtick={-8,-6,-4,-2,0,2,4,6,8},
      xticklabels={-8,-6,-4,-2,0,2,4,6,8},	
      xlabel={ \Huge{x [mm]}},
      xlabel near ticks,
      xmin=-8,
      xmax=5,
      ytick={0,1,2,3,4,5,6,7,8,9,10,15,20},	
      yticklabels={0,1,2,3,4,5,6,7,8,9,10,15,20},	
      ylabel={ \Huge{y [mm]}},
      ticklabel style = {font=\Huge},
      ylabel near ticks,
      ymin=0,
      ymax=8,
      ]

      \addplot[black, line width=2mm] table[x index=0, y index=1] {images/Data/OriginalConcave.dat};
      \addplot[black, line width=2mm, forget plot] table[x index=0, y index=1] {images/Data/OriginalConvex.dat};
      \draw [fill=gray, gray] (axis cs:-1.5,0) rectangle (axis cs:-1.0,3);
      \node (Omega_beam) at (axis cs:0.5, 0.5) [anchor=south] {\Huge{$\Omega_\mathrm{beam}$}};
      \node (in_beam) at (axis cs: -1.5, 2) {};
      \draw [thick] (Omega_beam) -- (in_beam);
      \node (Omega_p1) at (axis cs:-6, 1) [anchor=south] {\Huge{$\Omega_\mathrm{p}^{(1)}$}};
      \node (air_gap) at (axis cs:-5, 6) [anchor=south] {\Huge{$\Omega_\mathrm{p}^{(2)} = \Omega_\mathrm{air}$}};
      \node (Omega_p2) at (axis cs:2, 6) [anchor=south] {\Huge{$\Omega_\mathrm{p}^{(3)}$}};
  \end{axis}
\end{tikzpicture} 
\label{fig:zoomin}}
\caption{Left: 3D model of one-half of a Rabi-type Stern-Gerlach magnet. Right: Zoom in the magnet's pole region. Modified pictures from \cite{PelsIEEE}.}
\label{fig:stern-gerlach-model}
\end{figure}

We consider a real-world application, in particular a Rabi-type Stern-Gerlach magnet (see Figure \ref{fig:3dmodel}), similar to the one described in \cite{MasschaeleIEEE} and further studied in \cite{PelsIEEE, RoemerJCP}. 
This magnet is currently in use at KU Leuven. 
Stern-Gerlach magnets are typically employed for the magnetic separation of atom beams or clusters.
A key design requirement is a homogeneous magnetic field with a strong gradient. 
Due to design and manufacturing imperfections, the pole region might suffer from geometrical uncertainties, which in their turn affect the field homogeneity and gradient. 

The aim of this study is to apply the adaptive stochastic collocation method in order to quantify the impact of geometrical uncertainties onto the average magnetic field gradient in the magnet's beam area, which is the considered \gls{qoi}.
A further goal is to derive sufficiently accurate surrogate models, which can reliably replace the original model for computationally demanding \gls{uq} tasks, such as sensitivity analyses.

\label{sec:sterngerlachmodel}

All computations are performed using a linearized two-dimensional model of the magnet's cross-section, as in \cite{PelsIEEE}.
The magnet's pole region, denoted with $\Omega_\text{p}$, is the only domain which is spatially resolved.
Domain $\Omega_{\text{p}}$ is decomposed into distinct patches $\Omega_{\text{p}}^{(i)}$, $i=1,\ldots,3$, with numbering from left to right, such that $\Omega_{\text{p}} = \Omega_\text{p}^{(1)} \cup \Omega_\text{p}^{(2)} \cup \Omega_\text{p}^{(3)}$. 
Region $\Omega_\text{p}^{(2)}$ refers to the air gap inside the magnet's pole region, while regions $\Omega_\text{p}^{(1)}$ and $\Omega_\text{p}^{(3)}$ to the regions on the left and right of the air gap, respectively, as in Figure \ref{fig:zoomin}.
The contributions of the remaining yoke part and the coils are taken into account by a field-circuit coupling and a magnetic equivalent circuit \cite{PelsIEEE}. 
More precisely, in a first step, the magnetic vector potential and the magnetic flux through the iron yoke are computed for the entire geometry. 
The values are denoted as $A_z^0$ and $\Phi^0$, respectively. 
Then, the coupling is realized by imposing 
\begin{equation}
A_z = A_\Phi = \Phi/\Phi^0 A_z^0, \quad \ \text{on} \ \partial \Omega_\text{p},
\end{equation}
where $\Phi$ is recomputed for a different geometry using magnetic circuit theory. 
Let $I$ and $N_c$ represent the coil current and the number of turns in the winding, respectively. 
Then this relation can be abstractly written as $F(A_z,\Phi) = N_c I$, where $F$ refers to the magnetomotive force. 
For details, the reader is referred to \cite{PelsIEEE}. 
In summary, the field-circuit coupled problem reads
\begin{subequations}
\begin{align}
\text{div} \left(\nu \ \text{grad} A_z \right) &= 0, \quad \text{in} \ \Omega_\text{p}, \\
A_z -A_\Phi&= 0, \quad \text{on} \ \partial \Omega_\text{p}, \\
F(A_z,\Phi) &= N_c I,
\end{align}
\label{eq:mqs}%
\end{subequations}
where $\nu$ refers to the magnetic reluctivity. 
The magnetic flux density $\vec{B}$ is obtained as $\vec{B} = (\partial_y A_z,-\partial_x A_z,0)$. 
The magnet's beam area is denoted with $\Omega_\text{beam}$ and lies inside the air gap of the pole region, denoted with $\Omega_\text{air}$, where $\Omega_\text{air} = \Omega_\mathrm{p}^{(2)}$ (see Figure \ref{fig:zoomin}).
Denoting with $\tau\left(x, y \right) = \frac{\partial | \vec{B} | }{\partial x}$ the magnetic field gradient in the $x$-direction, the average field gradient in the beam area is given by
\begin{equation}
  \tau_{\text{avg}} = \frac{1}{|\Omega_{\text{beam}}|} \int_{\Omega_{\text{beam}}} \tau\left(x, y\right) \mathrm{d}\Omega.
\end{equation}
We note that $\tau_{\text{avg}}$ is here the \gls{qoi}. 

Isogeometric analysis (IGA) is employed for the spatial discretization \cite{hughes2005isogeometric}. 
In IGA, both the solution variable $A_z$ and the geometry are described in terms of non-uniform rational B-splines (NURBS). 
A generic NURBS curve reads 
\begin{equation}
\label{eq:nurbscurve}
\vec{R}(\xi) = \sum_{i=1}^N \vec{P}_i N_{i}^p(\xi), \quad \xi \in [0,1],
\end{equation}
where $\vec{P}_i$ and $N_i^p$ refer to a control point and a NURBS basis function of degree $p$, respectively. 
NURBS basis functions are defined as 
\begin{equation}
\label{eq:nurbsbasis}
N_i^p(\xi) = \frac{w_i B_i^p(\xi)}{\sum_{j=1}^N w_j B_j^p(\xi)},
\end{equation}
with weights $w_i$ and B-spline basis functions $B_i^p$, respectively.

The solid, black curves in Figure~\ref{fig:zoomin} correspond to the original NURBS curves defining the three patches $\Omega_\text{p}^{(i)}$, $i=1,2,3$, i.e. to the nominal geometry of the magnet's pole region.
We introduce random geometry deformations in the pole area by regarding the control points and weights of the NURBS curves as uncertain. 
More precisely, we introduce a total of $14$ \glspl{rv}, where $10$ \glspl{rv} correspond to the $x$ and $y$ coordinates of $5$ control points, while $4$ \glspl{rv} correspond to $4$ weights.
The nominal parameter values, $y_n^{\text{nom}}$, referring to the nominal pole geometry, are reported in Table~\ref{tab:stern-gerlach-nom}.
Due to lack of information regarding geometrical variations in the pole region, besides the accuracy limits of the manufacturing process, we only consider uniform distributions.
The limits of the uniform distributions are chosen such that the validity of the magnet pole's geometry is not violated.
The realizations of all uncertain parameters are given by $y_n = y_n^{\text{nom}} + Y_n(\theta)$, where $Y_n \sim \mathcal{U}\left(-1,1\right)$ for the coordinates and $Y_n \sim \mathcal{U}\left(0,1\right)$ for the weights.
With this modeling approach, all coordinate parameters are allowed a maximum deviation of $1$~mm, while the random weights introduce curve variations.

\begin{table}[b!]
\caption{Nominal parameter values for the Stern-Gerlach magnet.}
\centering\begin{tabular}{|c|c|c|}
\hline
\textbf{Parameter}  & \textbf{Nominal Value} & \textbf{Units} \\
\hline
   $x_1$ & $-2.38$ & mm\\
\hline 
  $y_1$ & $6.96$ & mm\\
\hline
  $x_2$ & $-2.38$ & mm\\
\hline 
  $y_2$ & $4.96$ & mm\\
\hline
  $x_3$ & $17.0$ & mm\\
\hline 
  $y_3$ & $20.0$ & mm\\
\hline
  $x_4$ & $-17.0$ & mm\\
\hline 
  $y_4$ & $20.0$ & mm\\
\hline
  $x_5$ & $-6.0$ & mm\\
\hline 
  $y_5$ & $4.0$ & mm\\
\hline
  $w_1$ & $0.85$ & --\\
\hline 
  $w_2$ & $0.85$ & --\\
\hline
  $w_3$ & $0.87$ & --\\
\hline 
  $w_4$ & $0.87$ & --\\
\hline
\end{tabular}
\label{tab:stern-gerlach-nom}
\end{table}

Then, we obtain a random reluctivity as 
\begin{equation}
\nu(\mathbf{y}) = \nu_\text{iron} \mathbbm{1}_{\Omega_{\text{p}}^{(1)}(\mathbf{y})} + \nu_\text{air} \mathbbm{1}_{\Omega_{\text{p}}^{(2)}(\mathbf{y})} + \nu_\text{iron} \mathbbm{1}_{\Omega_{\text{p}}^{(3)}(\mathbf{y})}, 
\end{equation}
with $\mathbbm{1}_{\Omega_{\text{p}}^{(i)}}$ denoting the characteristic function of patch $i$ and $\nu_\text{iron}$ and $\nu_\text{air}$ denoting the reluctivity of iron and air, respectively. 
Accordingly, the parametric field-circuit coupled problem reads 
\begin{subequations}
\begin{align}
\text{div} \left(\nu(\mathbf{y}) \ \text{grad} A_z \right) &= 0, \quad \text{in} \ \Omega_\text{p},  \\ 
A_z - A_\Phi &= 0, \quad \text{on} \ \partial \Omega_\text{p}, \\
F(A_z, \Phi) &= N_c I, 
\end{align}
\label{eq:mqs_parametric}%
\end{subequations}
$\rho_\mathbf{Y}$-almost everywhere in $\Gamma$. Assuming that problem \eqref{eq:mqs_parametric} is well-posed, the \gls{qoi} $\tau_{\text{avg}}$ is itself a random variable.

\subsubsection{Surrogate model accuracy}
\label{sec:sterngerlach-accuracy}
We employ the dimension-adaptive Algorithm~\ref{algo:gensmolyak}, based on both Clenshaw-Curtis and Leja nodes, in order the approximate the $14$-dimensional parametric Stern-Gerlach magnet model.
The approximation accuracy is measured with the cross-validation error metrics given in \eqref{eq:cverr_max}, \eqref{eq:cverr_mean} and \eqref{eq:cverr_rms}, based on a random sample of size $M = 10^4$.
The results are presented in Figure~\ref{fig:sterngerlach_cverrz}.
As can be observed, both rules have an equivalent performance, reaching similar accuracies for approximately the same number of model evaluations.

\begin{figure}[t!]
\centering
   \begin{tikzpicture} 
      \begin{semilogyaxis}[width=0.5\textwidth, xlabel=Model evaluations, ylabel=$\epsilon_{\text{cv}}$, legend pos=outer north east, legend style={fill=none}]
	\addplot[mark=None, gray, thick] table[x index=0, y index=1, col sep=comma]{plot_data/sterngerlach/SternGerlach_adaptCC_1e4.txt};
	\addplot[mark=None, gray, thick, dotted] table[x index=0, y index=2, col sep=comma]{plot_data/sterngerlach/SternGerlach_adaptCC_1e4.txt};
	\addplot[mark=None, gray, thick, dashed] table[x index=0, y index=3, col sep=comma]{plot_data/sterngerlach/SternGerlach_adaptCC_1e4.txt};
	\addplot[mark=None, black, thick] table[x index=0, y index=1]{plot_data/sterngerlach/SternGerlach_dali_1e4.txt};
	\addplot[mark=None, black, thick, dotted] table[x index=0, y index=2]{plot_data/sterngerlach/SternGerlach_dali_1e4.txt};
	\addplot[mark=None, black, thick, dashed] table[x index=0, y index=3]{plot_data/sterngerlach/SternGerlach_dali_1e4.txt};
	\legend{{Clenshaw-Curtis, $\epsilon_{\text{cv}, \text{max}}$}, {Clenshaw-Curtis, $\epsilon_{\text{cv}, \text{mean}}$}, {Clenshaw-Curtis, $\epsilon_{\text{cv}, \text{RMS}}$}, {Leja, $\epsilon_{\text{cv}, \text{max}}$}, {Leja, $\epsilon_{\text{cv}, \text{mean}}$}, {Leja, $\epsilon_{\text{cv}, \text{RMS}}$}}
      \end{semilogyaxis}
    \end{tikzpicture}
\caption{Cross-validation errors for the Stern-Gerlach magnet with 14 uniform random inputs. The approximations are based on Clenshaw-Curtis and uniform (unweighted) Leja rules. A validation set with $10^4$ parameter realizations is used.}
\label{fig:sterngerlach_cverrz}
\end{figure}

\subsubsection{Moment computations}
\label{sec:sterngerlach-moments}
We employ the randomly generated cross-validation sample used in Section~\ref{sec:sterngerlach-accuracy} to compute the expected value and the variance of $\tau_{\text{avg}}$ via \gls{mc} sampling.
The corresponding \gls{mc} results, along with \gls{mc}'s \gls{rms} error, $\epsilon_{\text{MC}, \text{RMS}}$, and normalized \gls{rms} error, $\epsilon_{\text{MC}, \text{NRMS}}$, are 
\begin{align*}
 \mathbb{E}\left[\tau_{\text{avg}}\right] &= -238.1529~\mathrm{T/m}, \\
 \mathbb{V}\left[\tau_{\text{avg}}\right] &= 21.7974~\mathrm{T^2/m^2}, \\
 \epsilon_{\text{MC}, \text{RMS}} &\approx 0.05~\mathrm{T/m}, \\
 \epsilon_{\text{MC}, \text{NRMS}} &\approx 2 \cdot 10^{-4}. 
\end{align*}
The \gls{mc} results given above are used as reference values in the computations of relative errors with both dimension-adaptive approaches, i.e. Clenshaw-Curtis and Leja-based.
The relative errors with respect to the expected value and the variance of the \gls{qoi} are presented in Figure~\ref{fig:sterngerlach_relerrz}.

Regarding the expected value, both approaches converge to a relative error almost identical to $\epsilon_{\text{MC}, \text{NRMS}}$, as shown in Figure~\ref{fig:sterngerlach_mean}.
After only 300 model evaluations, both dimension-adaptive approaches reach the same accuracy provided by the sampling-based approach with $10^4$ random samples.
Both approaches converge to the same relative errors also in the case of the variance, as shown in Figure~\ref{fig:sterngerlach_var}.
In that case, absolute convergence is observed after less than 2000 model evaluations for the Clenshaw-Curtis-based collocation and after approximately 4000 evaluations for the Leja-based collocation.
However, the stagnation in the relative error can already be observed for both methods after approximately 1000 model evaluations.
Again, both methods can be regarded as equivalent, since no significant differences are observed. 

\begin{figure}[t!]
  \subfigure[Relative errors for the expected value. \label{fig:sterngerlach_mean}]{
  \begin{tikzpicture}
      \begin{semilogyaxis}[width=0.45\textwidth, xlabel=Model evaluations, ylabel=$\epsilon_{\text{rel}}$, legend pos=north east]
	\addplot[mark=None, gray, thick] table[x index=0, y index=1, skip coords between index={15}{59}]{plot_data/sterngerlach/SternGerlach_adaptCC_rel_errz.txt};
	\addplot[mark=None, black, thick] table[x index=0, y index=1, skip coords between index={15}{59}]{plot_data/sterngerlach/SternGerlach_dali_rel_errz.txt};
	\legend{Clenshaw-Curtis, Leja}
      \end{semilogyaxis}
  \end{tikzpicture}
  }
  \subfigure[Relative errors for the variance. \label{fig:sterngerlach_var}]{
  \begin{tikzpicture}
      \begin{semilogyaxis}[width=0.45\textwidth, xlabel=Model evaluations, legend pos=north east]
	\addplot[mark=None, gray, thick] table[x index=0, y index=2]{plot_data/sterngerlach/SternGerlach_adaptCC_rel_errz.txt};
	\addplot[mark=None, black, thick] table[x index=0, y index=2]{plot_data/sterngerlach/SternGerlach_dali_rel_errz.txt};	\legend{Clenshaw-Curtis, Leja}
      \end{semilogyaxis}
  \end{tikzpicture}
  }
\caption{Moment relative errors for the Stern-Gerlach magnet with 14 uniform random inputs. The approximations are based on Clenshaw-Curtis and uniform (unweighted) Leja rules. The reference moment values are computed using Monte Carlo sampling with $10^4$ samples.}
\label{fig:sterngerlach_relerrz}
\end{figure}

\subsubsection{Surrogate-based sensitivity analysis}
\label{sec:sterngerlach-sa}
As a final numerical experiment, we perform a surrogate-based Sobol sensitivity analysis \cite{sobol2001}.
To that end, we employ the two approaches discussed in Section~\ref{sec:quad_postproc}, namely the sampling-based algorithm from \cite{saltelli2002} and the change-of-basis method suggested in \cite{buzzard2013, porta2014}.

In the former case, we opt for an input sample of size $M=10^4$, thus requiring $3 \cdot 10^5$ model evaluations.
On a standard desktop and assuming no use of parallel computing resources, this number of evaluations can be executed in 2-3 minutes, using a surrogate model with $5000$ terms.
Using the original model would result in more than $100$ days of computation, on the same machine.
The change-of-basis method is even more efficient, yielding the sensitivity results in less than $3$ seconds, again considering a surrogate model with $5000$ terms.
In both cases the cost of the sensitivity analysis can be regarded as negligible next to the costs for constructing the surrogate model.

\begin{figure}[t!]
 \centering
 \begin{tikzpicture}
    \begin{axis}[ybar, symbolic x coords={$w_1$, $w_2$, $w_3$, $w_4$, $x_1$, $x_2$}, xtick=data, ylabel={First-order Sobol index}, xlabel={Parameters}, legend pos=north east, ytick={0.05, 0.1, 0.15, 0.2, 0.25, 0.3}, tick label style={/pgf/number format/fixed}]
    \addplot[black, fill=black] coordinates {($w_1$, 0.23)  ($w_2$, 0.28) ($w_3$, 0.05) ($w_4$, 0.16) ($x_1$, 0.07) ($x_2$, 0.21)};
    \addplot[gray, fill=gray] coordinates {($w_1$, 0.23)  ($w_2$, 0.27) ($w_3$, 0.05) ($w_4$, 0.16) ($x_1$, 0.06) ($x_2$, 0.21)};
    \legend{Sampling, Basis change}
    \end{axis}
  \end{tikzpicture}
  \caption{Surrogate-based sensitivity analysis results for the Stern-Gerlach magnet with 14 uniform random inputs. Only first-order Sobol indices are shown. The parameters with Sobol indices smaller that 0.01 have been omitted.}
\label{fig:sterngerlach_sobol}
\end{figure}
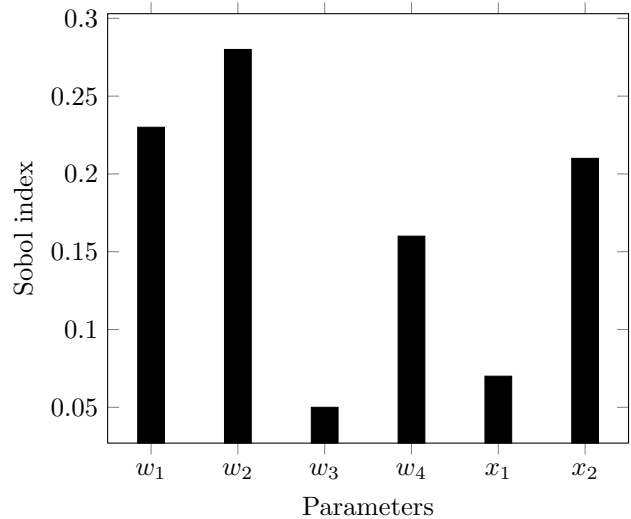

The sensitivity analysis results are presented in Figure~\ref{fig:sterngerlach_sobol}.
We only present results corresponding to a Clenshaw-Curtis-based surrogate model, however, almost identical results have been obtained with the Leja rule.
We omit all parameters with contributions below $1\%$, i.e. with Sobol indices smaller than $0.01$.
The sum of all first-order Sobol indices is equal to $0.997$ using the sampling-based approach and equal to $0.98$ using the change-of-basis approach.
Therefore, we only present first-order Sobol indices, since sensitivities of higher order may be safely omitted.

With the exception of an $1\%$ difference observed for parameters $w_2$ and $x_1$, both approaches yield almost identical results.
We note that the sensitivity results obtained with the change-of-basis approach coincide with the ones derived after an adaptive generalized polynomial chaos approach, similar to the one presented in \cite{conf/enumath/Migliorati13}, i.e. by directly constructing a Legendre-chaos approximation without any basis transformation.
Considering the sampling-based sensitivity analysis and its good agreement with the other two approaches, we may conclude that the surrogate model has reliably substituted the original one in this study.

As can be observed, only 6 out of the initially considered 14 parameters seem to have a significant influence on the \gls{qoi}.
In the light of this result, the input dimensionality of the parametrized model can be significantly reduced.
The results also indicate that shape variations, given by the random weights $w_i$, $i=1,\dots,4$, play a much more important role than coordinate shifts, which should be taken into account in a further attempt to improve the magnet model.

\section{Summary and Conclusions}
\label{sec:conclusions}

In this work we have employed the stochastic collocation method for \gls{uq} in \gls{emf} problems with bounded random inputs.
Both uniform and non-uniform random inputs have been considered.
In the latter case, we have employed positively skewed beta distributions as a typical example.
A dimension-adaptive algorithm based on nested univariate collocation points has been employed in all multivariate cases.
Two families of nested collocation points, provided by the Clenshaw-Curtis and Leja rules, have been examined in terms of interpolation and quadrature accuracy.

In the case of uniform input distributions, the Leja rule is found to be advantageous compared to the Clenshaw-Curtis rule in terms of interpolation accuracy, considering an analytical, academic model with a very smooth input-to-output map. 
For the same model, the Clenshaw-Curtis rule outperforms the Leja rule in terms of quadrature accuracy, as measured in statistical moment computations.
No obvious advantages can be observed for either choice of collocation points in the numerical studies concerning the real-world \gls{emf} application, where both rules can be regarded as comparable.

Interesting results are reported in the case of beta-distributed inputs.
In the univariate case, the Clenshaw-Curtis rule is found to be advantageous in terms of worst-case performance, but inferior in terms of average performance.
A similar result is obtained when comparing unweighted and weighted Leja rules.
This result indicates that surrogate models based on uniform input distributions should probably be preferred if worst-case performance is of main interest.
However, the same result is not observed in the multivariate case, where the weighted Leja rule outperforms both the Clenshaw-Curtis and the unweighted Leja rules in all metrics, i.e. in terms of both average and worst-case performance.
Finally, considering moment estimations, the weighted Leja rule is found to outperform the Clenshaw-Curtis rule with modified weights in all occasions, i.e. for all considered moments and in both univariate and multivariate cases.

Based on the results of the numerical experiments presented in this paper, we may conclude that Leja rules present a reliable choice of nested collocation points, producing accurate results both in the context of interpolation and quadrature.
The versatility of Leja points allows the rule to be employed for arbitrary input \glspl{pdf}, however, its performance against competitive rules in a wider variety of cases should be further examined.
An extension of this work should consider cases where dedicated nested collocation points exist, e.g. truncated normal \cite{burkardt2014} or other distributions.

\subsubsection*{Acknowledgments}
All authors would like to thank our colleague, Andreas Pels, for providing the Stern-Gerlach magnet model.
We would also like to thank the two anonymous reviewers for their comments and suggestions.
The first and third author would like to acknowledge the support of the Graduate School of Computational Engineering, Technische Universit\"{a}t Darmstadt. 

% Bibliography
\bibliographystyle{siam}
\bibliography{references}

\begin{thebibliography}{10}

\bibitem{journals/siamrev/BabuskaNT10}
{\sc I.~Babuska, F.~Nobile, and R.~Tempone}, {\em {A Stochastic Collocation
  Method for Elliptic Partial Differential Equations with Random Input Data}},
  SIAM Review,  (2010), pp.~317--355.

\bibitem{journals/siamnum/BabuskaTZ04}
{\sc I.~Babuska, R.~Tempone, and G.~E. Zouraris}, {\em {Galerkin Finite Element
  Approximations of Stochastic Elliptic Partial Differential Equations}}, SIAM
  Journal on Numerical Analysis, 42 (2004), pp.~800--825.

\bibitem{journals/adcm/BarthelmannNR00}
{\sc V.~Barthelmann, E.~Novak, and K.~Ritter}, {\em {High Dimensional
  Polynomial Interpolation on Sparse Grids}}, Advances in Computational
  Mathematics, 12 (2000), pp.~273--288.

\bibitem{Bellman:57}
{\sc D.~Bellman}, {\em {Dynamic Programming}}, Princeton University Press,
  1957.

\bibitem{berrut2004barycentric}
{\sc J.~Berrut and L.~Trefethen}, {\em {Barycentric Lagrange Interpolation}},
  SIAM Review, 46 (2004), pp.~501--517.

\bibitem{journals/jcphy/BlatmanS11}
{\sc G.~Blatman and B.~Sudret}, {\em {Adaptive Sparse Polynomial Chaos
  Expansion Based on Least Angle Regression}}, Journal of Computational
  Physics, 230 (2011), pp.~2345--2367.

\bibitem{bungartz_griebel_2004}
{\sc H.-J. Bungartz and M.~Griebel}, {\em {Sparse Grids}}, Acta Numerica, 13
  (2004), pp.~147--269.

\bibitem{burkardt2014}
{\sc J.~Burkardt}, {\em {The Truncated Normal Distribution}}, 2014.
\newblock Department of Scientific Computing, Florida State University.

\bibitem{buzzard2013}
{\sc G.~T. Buzzard}, {\em {Efficient Basis Change for Sparse-Grid Interpolating
  Polynomials with Application to T-Cell Sensitivity Analysis}}, Computational
  Biology Journal,  (2013).

\bibitem{caflisch1998}
{\sc R.~E. Caflisch}, {\em {Monte Carlo and Quasi-Monte Carlo Methods}}, Acta
  Numerica, 7 (1998), p.~1–49.

\bibitem{journals/focm/ChkifaCS14}
{\sc A.~Chkifa, A.~Cohen, and C.~Schwab}, {\em {High-Dimensional Adaptive
  Sparse Polynomial Interpolation and Applications to Parametric PDEs}},
  Foundations of Computational Mathematics, 14 (2014), pp.~601--633.

\bibitem{clenshaw1960}
{\sc C.~W. Clenshaw and A.~R. Curtis}, {\em {A Method for Numerical Integration
  on an Automatic Computer}}, Numerische Mathematik, 2 (1960), pp.~197--205.

\bibitem{eigel2014}
{\sc M.~Eigel, C.~J. Gittelson, C.~Schwab, and E.~Zander}, {\em {Adaptive
  Stochastic Galerkin FEM}}, Computer Methods in Applied Mechanics and
  Engineering, 270 (2014), pp.~247 -- 269.

\bibitem{Eldred2009comparison}
{\sc M.~S. Eldred}, {\em {Comparison of Non-Intrusive Polynomial Chaos and
  Stochastic Collocation Methods for Uncertainty Quantification}}, In: AIAA,
  (2009).

\bibitem{journals/jocs/FeinbergL15}
{\sc J.~Feinberg and H.~P. Langtangen}, {\em {Chaospy: An Open Source Tool for
  Designing Methods of Uncertainty Quantification}}, Journal of Computational
  Science, 11 (2015), pp.~46--57.

\bibitem{journals/computing/GerstnerG03}
{\sc T.~Gerstner and M.~Griebel}, {\em {Dimension-Adaptive Tensor-Product
  Quadrature}}, Computing, 71 (2003), pp.~65--87.

\bibitem{Ghanem1991}
{\sc R.~G. Ghanem and P.~D. Spanos}, {\em Stochastic Finite Elements: A
  Spectral Approach}, Springer-Verlag New York, Inc., New York, NY, USA, 1991.

\bibitem{giles2015}
{\sc M.~B. Giles}, {\em {Multilevel Monte Carlo Methods}}, Acta Numerica, 24
  (2015), p.~259–328.

\bibitem{journals/siamsc/GiraldiLLMN14}
{\sc L.~Giraldi, A.~Litvinenko, D.~Liu, H.~G. Matthies, and A.~Nouy}, {\em {To
  Be or Not to Be Intrusive? The Solution of Parametric and Stochastic
  Equations - the "Plain Vanilla" Galerkin Case}}, SIAM Journal on Scientific
  Computing, 36 (2014).

\bibitem{hackbusch2014}
{\sc W.~Hackbusch}, {\em {Numerical Tensor Calculus}}, Acta Numerica, 23
  (2014), p.~651–742.

\bibitem{hughes2005isogeometric}
{\sc T.~J. Hughes, J.~A. Cottrell, and Y.~Bazilevs}, {\em {Isogeometric
  Analysis: CAD, Finite Elements, NURBS, Exact Geometry and Mesh Refinement}},
  Computer Methods in Applied Mechanics and Engineering, 194 (2005),
  pp.~4135--4195.

\bibitem{journals/jcphy/JakemanW15}
{\sc J.~D. Jakeman and T.~Wildey}, {\em {Enhancing Adaptive Sparse Grid
  Approximations and Improving Refinement Strategies Using Adjoint-Based A
  Posteriori Error Estimates}}, Journal of Computational Physics, 280 (2015),
  pp.~54--71.

\bibitem{jantsch2018}
{\sc P.~Jantsch, C.~G. Webster, and G.~Zhang}, {\em {On the Lebesgue Constant
  of Weighted Leja Points for Lagrange Interpolation on Unbounded Domains}},
  IMA Journal of Numerical Analysis,  (2018).

\bibitem{jin2015finite}
{\sc J.-M. Jin}, {\em The Finite Element Method in Electromagnetics}, John
  Wiley \& Sons, 2015.

\bibitem{journals/toms/KlimkeW05}
{\sc A.~Klimke and B.~I. Wohlmuth}, {\em {Algorithm 847: Spinterp: Piecewise
  Multilinear Hierarchical Sparse Grid Interpolation in MATLAB}}, ACM
  Transactions on Mathematical Software, 31 (2005), pp.~561--579.

\bibitem{leja1957}
{\sc F.~Leja}, {\em Sur certaines suites liées aux ensembles plans et leur
  application à la représentation conforme}, Annales Polonici Mathematici, 4
  (1957), pp.~8--13.

\bibitem{loukrezis2018}
{\sc D.~Loukrezis}, {\em {Dimension Adaptive Leja Interpolation (DALI)}}.
\newblock URL: https://github.com/dlouk/DALI. Accessed: 15.09.2018.

\bibitem{MasschaeleIEEE}
{\sc B.~Masschaele, T.~Roggen, H.~{De Gersem}, E.~Janssens, and T.~T. Nguyen},
  {\em {Design of a Strong Gradient Magnet for the Deflection of
  Nanoclusters}}, IEEE Transactions on Applied Superconductivity, 22 (2012).

\bibitem{MatthiesGalerkin}
{\sc H.~G. Matthies and A.~Keese}, {\em {Galerkin Methods for Linear and
  Nonlinear Elliptic Stochastic Partial Differential Equations}},
  Informatik-Berichte der Technischen Universit{\"a}t Braunschweig, 2003-08.

\bibitem{conf/enumath/Migliorati13}
{\sc G.~Migliorati}, {\em {Adaptive Polynomial Approximation by Means of Random
  Discrete Least Squares}}, in {ENUMATH}, A.~Abdulle, S.~Deparis, D.~Kressner,
  F.~Nobile, and M.~Picasso, eds., vol.~103 of {Lecture Notes in Computational
  Science and Engineering}, Springer, 2013, pp.~547--554.

\bibitem{journals/siamsc/MiglioratiNST13}
{\sc G.~Migliorati, F.~Nobile, E.~von Schwerin, and R.~Tempone}, {\em
  {Approximation of Quantities of Interest in Stochastic PDEs by the Random
  Discrete L2 Projection on Polynomial Spaces}}, SIAM Journal on Scientific
  Computing, 35 (2013).

\bibitem{journals/focm/MiglioratiNST14}
\leavevmode\vrule height 2pt depth -1.6pt width 23pt, {\em {Analysis of
  Discrete L2 Projection on Polynomial Spaces with Random Evaluations}},
  Foundations of Computational Mathematics, 14 (2014), pp.~419--456.

\bibitem{journals/siamsc/NarayanJ14}
{\sc A.~Narayan and J.~D. Jakeman}, {\em {Adaptive Leja Sparse Grid
  Constructions for Stochastic Collocation and High-Dimensional
  Approximation}}, SIAM Journal on Scientific Computing, 36 (2014).

\bibitem{Nobile2015}
{\sc F.~Nobile, L.~Tamellini, and R.~Tempone}, {\em {Comparison of
  Clenshaw--Curtis and Leja Quasi-Optimal Sparse Grids for the Approximation of
  Random PDEs}}, in Spectral and High Order Methods for Partial Differential
  Equations ICOSAHOM 2014: Selected papers from the ICOSAHOM conference, June
  23-27, 2014, Salt Lake City, Utah, USA, R.~M. Kirby, M.~Berzins, and J.~S.
  Hesthaven, eds., Springer International Publishing, Cham, 2015, pp.~475--482.

\bibitem{journals/siamnum/NobileTW08}
{\sc F.~Nobile, R.~Tempone, and C.~G. Webster}, {\em {A Sparse Grid Stochastic
  Collocation Method for Partial Differential Equations with Random Input
  Data}}, SIAM Journal on Numerical Analysis, 46 (2008), pp.~2309--2345.

\bibitem{journals/siamnum/NobileTW08a}
\leavevmode\vrule height 2pt depth -1.6pt width 23pt, {\em {An Anisotropic
  Sparse Grid Stochastic Collocation Method for Partial Differential Equations
  with Random Input Data}}, SIAM Journal on Numerical Analysis, 46 (2008),
  pp.~2411--2442.

\bibitem{PelsIEEE}
{\sc A.~Pels, Z.~Bontinck, J.~Corno, H.~{De Gersem}, and S.~Sch{\"o}ps}, {\em
  {Optimization of a Stern-Gerlach Magnet by Magnetic Field-Circuit Coupling
  and Isogeometric Analysis}}, IEEE Transactions on Magnetics, 51 (2015).

\bibitem{porta2014}
{\sc G.~Porta, L.~Tamellini, V.~Lever, and M.~Riva}, {\em {Inverse Modeling of
  Geochemical and Mechanical Compaction in Sedimentary Basins through
  Polynomial Chaos Expansion}}, Water Resources Research, 50 (2014),
  pp.~9414--9431.

\bibitem{RoemerJCP}
{\sc U.~R{\"o}mer, S.~Sch{\"o}ps, and H.~De~Gersem}, {\em {A Defect Corrected
  Finite Element Approach for the Accurate Evaluation of Magnetic Fields on
  Unstructured Grids}}, Journal of Computational Physics, 335 (2017),
  pp.~688--699.

\bibitem{saltelli2002}
{\sc A.~Saltelli}, {\em {Making Best Use of Model Evaluations to Compute
  Sensitivity Indices}}, Computer Physics Communications, 145 (2002),
  pp.~280--297.

\bibitem{SchiechePhD}
{\sc B.~Schieche}, {\em {Unsteady Adaptive Stochastic Collocation on Sparse
  Grids}}, PhD Thesis, TU Darmstadt, 2012.

\bibitem{Smolyak1963}
{\sc S.~A. Smolyak}, {\em {Quadrature and Interpolation Formulas for Tensor
  Products of Certain Classes of Functions}}, Doklady Akademii Nauk SSSR, 4
  (1963), pp.~240--243.

\bibitem{sobol2001}
{\sc I.~M. Sobol}, {\em {Global Sensitivity Indices for Nonlinear Mathematical
  Models and their Monte Carlo Estimates}}, Mathematics and Computers in
  Simulation, 55 (2001), pp.~271--280.

\bibitem{sommariva2013fast}
{\sc A.~Sommariva}, {\em {Fast Construction of Fej{\'e}r and Clenshaw--Curtis
  Rules for General Weight Functions}}, Computers \& Mathematics with
  Applications, 65 (2013), pp.~682--693.

\bibitem{sudret2008}
{\sc B.~Sudret}, {\em {Global Sensitivity Analysis using Polynomial Chaos
  Expansions}}, Reliability Engineering \& System Safety, 93 (2008), pp.~964 --
  979.
\newblock Bayesian Networks in Dependability.

\bibitem{csqi}
{\sc L.~Tamellini and F.~Nobile}, {\em {Sparse Grids MATLAB Kit}}.
\newblock URL: http://csqi.epfl.ch. Accessed: 15.09.2018.

\bibitem{trefethen2008}
{\sc L.~N. Trefethen}, {\em {Is Gauss Quadrature Better than
  Clenshaw–Curtis?}}, SIAM Review, 50 (2008), pp.~67--87.

\bibitem{UschmajewPhD}
{\sc A.~Uschmajew}, {\em {Zur Theorie der Niedrigrangapproximation in
  Tensorprodukten von Hilbertr{\"a}umen}}, PhD Thesis, TU Berlin, 2013.

\bibitem{xiu2010}
{\sc D.~Xiu}, {\em {Numerical Methods for Stochastic Computations: A Spectral
  Method Approach}}, Princeton University Press, Princeton, NJ, USA, 2010.

\bibitem{journals/siamsc/XiuH05}
{\sc D.~Xiu and J.~S. Hesthaven}, {\em {High-Order Collocation Methods for
  Differential Equations with Random Inputs}}, SIAM Journal on Scientific
  Computing, 27 (2005), pp.~1118--1139.

\bibitem{journals/siamsc/XiuK02}
{\sc D.~Xiu and G.~E. Karniadakis}, {\em {{The Wiener-Askey Polynomial Chaos
  for Stochastic Differential Equations}}}, SIAM Journal on Scientific
  Computing, 24 (2002), pp.~619--644.

\end{thebibliography}
\end{document}